\documentclass[twocolumn]{aastex63}
\usepackage{amsmath,amstext}
\usepackage[T1]{fontenc}
\usepackage{apjfonts} 
\usepackage{multirow}
\usepackage[figure,figure*]{hypcap}
\usepackage{booktabs,caption}
\usepackage{subcaption}
\usepackage{mwe}
\received{~--- xxx}
\revised{~--- xxx}
\accepted{~--- xxx}

\shorttitle{AUDF South Catalog}
\shortauthors{Saha et al.}

\begin{document}
\title{AstroSat UV Deep Field South -- I. Far and Near-ultraviolet Source Catalog of the GOODS South region}

\correspondingauthor{Kanak Saha}
\email{kanak@iucaa.in}


\author[0000-0002-8768-9298]{Kanak Saha}
\affiliation{Inter-University Centre for Astronomy and Astrophysics, Ganeshkhind, Post Bag 4, Pune 411007, India}

\author[0009-0003-8568-4850]{Soumil Maulick}
\affiliation{Inter-University Centre for Astronomy and Astrophysics, Ganeshkhind, Post Bag 4, Pune 411007, India}

\author[0009-0009-7497-3431]{Pushpak Pandey}
\affiliation{Inter-University Centre for Astronomy and Astrophysics, Ganeshkhind, Post Bag 4, Pune 411007, India}

\author[0000-0003-4594-6943]{Souradeep Bhattacharya}
\affiliation{Inter-University Centre for Astronomy and Astrophysics, Ganeshkhind, Post Bag 4, Pune 411007, India}

\author[0000-0002-2870-7716]{Anshuman Borgohain}
\affiliation{Department of Physics, Tezpur University, Napaam, India}

\author[0000-0003-4531-0945]{Chayan Mondal}
\affiliation{Inter-University Centre for Astronomy and Astrophysics, Ganeshkhind, Post Bag 4, Pune 411007, India}

\author[0000-0002-9946-4731]{Marc Rafelski}
\affiliation{Space Telescope Science Institute, 3700 San Martin Drive, Baltimore, MD 21218, USA}
\affiliation{Department of Physics and Astronomy, Johns Hopkins University, Baltimore, MD 21218, USA}

\author[0009-0008-2970-9845]{Manish Kataria}
\affiliation{Inter-University Centre for Astronomy and Astrophysics, Ganeshkhind, Post Bag 4, Pune 411007, India}

\author[0000-0002-7064-5424]{Harry I. Teplitz}
\affiliation{IPAC, Mail Code 314-6, California Institute of Technology, 1200 E. California Blvd., Pasadena CA, 91125, USA}

\author[0000-0001-6350-7421]{Shyam N. Tandon}
\affiliation{Inter-University Centre for Astronomy and Astrophysics, Ganeshkhind, Post Bag 4, Pune 411007, India}

\author[0000-0001-8156-6281]{Rogier A. Windhorst}
\affiliation{School of Earth \& Space Exploration, Arizona State University, Tempe, AZ 85287-1404, USA}

\author[0000-0002-1723-6330]{Bruce G. Elmegreen}
\affiliation{Katonah, NY 10536, USA}

\author[0000-0002-8505-4678]{Edmund Christian Herenz}
\affiliation{Inter-University Centre for Astronomy and Astrophysics, Ganeshkhind, Post Bag 4, Pune 411007, India}

\author[0000-0001-7016-5220]{Michael Rutkowski}
\affiliation{Minnesota State University-Mankato, Mankato, MN 56001, USA}



\begin{abstract}
We present the AstroSat UV Deep Field South (AUDFs), an imaging survey using the wide-field Ultraviolet Imaging Telescope on board AstroSat. AUDFs covers $\sim 236$ arcmin$^{2}$ of the sky area, including the Great Observatories Origins Deep Survey (GOODS) South field in F154W and N242W filters. The deep and shallow parts of AUDFs have exposure time $\sim 62000$ and $\sim31000$ sec respectively, in the F154W filter, while in the N242W filter, they are $\sim 64000$ and $\sim34000$ sec. These observations reached a $3\sigma$ depth of 27.2 and 27.7 AB mag with a $50\%$ completeness limit of 27 and 27.6 AB mag in the F154W and N242W filters, respectively. With the acquired depth, AUDFs is the deepest far and near-UV imaging data covering the largest area known to date at $1.2\arcsec - 1.6\arcsec$ spatial resolution. Two primary catalogs were constructed for the F154W and N242W filters, each containing 13495 and 19374 sources brighter than the 3$\sigma$ detection limit, respectively. Our galaxy counts power-law slope $\sim0.43$~dex~mag$^{-1}$ in the N242W filter matches well with HST/WFC3/UVIS observations.

A wide range of extra-galactic science can be achieved with this unique data, such as providing a sample of galaxies emitting ionizing photons in the redshift range $z \sim 1 - 3$ and beyond; constraining the UV luminosity function, investigating the extended-UV (XUV) emission around star-forming galaxies and UV morphologies for $z < 1$. The UV catalog will enhance the legacy value of the existing optical/IR imaging and spectroscopic observations from ground and space-based telescopes on the GOODS South field.

\end{abstract}


\section{Introduction} 
\label{sec:intro}
Deep fields are windows to discover the past history of galaxy formation in the universe \citep{Robertsoneatl2015,Robertson2022}. Typically, deep fields are created by long exposures of a particular patch of the sky, that is carefully chosen to be devoid of bright stars, in a particular wavelength range of interest to reveal fainter and distant objects \citep{Cowieetal1991}. As the exposure time increases, fainter objects with low surface brightness start popping up in the field. These fainter objects could either be intrinsically faint local galaxies e.g., the ultra-diffuse galaxies \citep{vanDokkumetal2015,Forbesetal2019,Iodiceetal2020}; clumpy features e.g., clumps beyond the optical boundary \citep{Thilkeretal2007, Elmegreenetal2009, Borgohainetal2022}; low-luminous dwarf satellites around bigger galaxies \citep{Carlstenetal2020} or distant galaxies that are apparently fainter due to the surface brightness dimming \citep{Tolman1930,HubbleTolman1935}. Multi-wavelength observations of these faint galaxies, from far-ultraviolet to infrared with high sensitivity and resolution, play a key role in deciphering our understanding of galaxy growth and evolution \citep{BabulFerguson1996,Vandenbergh2002,Robertson2022}.   

The Hubble Space Telescope (HST), with its highly sensitive Advanced Camera for Surveys (ACS) and Wide-Field Camera 3 (WFC3), has revolutionized the deep field exploration by producing high-resolution deep imaging of several patches on the sky e.g., the Hubble Ultra Deep Field (HUDF; \citet{beckwith2006, Ellisetal2013, koekemoer2011, Illingworthetal2013,teplitz2013}), the Great Observatories Origins Deep Survey (GOODS) South and North \citep{dickinson2003, giavalisco2004, Windhorstetal2011,teplitz2013}, the Hubble Frontier Fields clusters and Parallels \citep{Lotzetal2017,Shipleyetal2018,Paguletal2021}, the Hubble Deep Field South and North \citep{Fergusonetal2000, dickinson2003}, the Cosmic Evolution Survey (COSMOS; \cite{Scovilleetal2007}), the Cosmic Assembly Near-infrared Deep Extragalactic Legacy Survey (CANDELS; \cite{grogin2011,koekemoer2011}). 

In particular, the GOODS-South \citep{giavalisco2004}, containing the HUDF, has been visited by almost all the state-of-the-art telescopes from ground and space. Thus, a wealth of multi-wavelength imaging observations from ultraviolet to infrared and millimeter are available from HST, Galaxy Evolution Explorer (GALEX), Herschel, Spitzer, ALMA, Chandra, the Very Large Telescope (VLT) and now the James Webb Space Telescope (JWST) on the GOODS South field \citep{Retzlaffetal2010, Elbazetal2011,beckwith2006, Ashbyetal2015, martin2005, Mainierietal2005, Luoetal2017, Fontanaetal2014, Dunlopetal2017, Whittakeretal2019,Francoetal2020,Eisensteinetal2023}. Such multi-wavelength data accompanied by spectroscopic follow-up \citep{Momchevaetal2016,Baconetal2017,McLure18,Revalskietal2023} has not only proven to be a goldmine for identifying high-redshift galaxies from the early universe \citep{Bouwensetal2010}, and constraining their possible role on the cosmic reionization \citep{Robertsonetal2010,Bunkeretal2010,Robertsonetal2013,Robertson2022} but also provided us with an wealth of information on physical properties of galaxies; for example, colors and luminosity function (LF) of high-z galaxies \citep{bouwens2004, Dunlopetal2013, Oeschetal2013, Ellisetal2013,Bouwensetal2015, Alavietal2016}, morphology of distant galaxies \citep{Elmegreenetal2005b, straughnetal2006, EE2014}, star-formation rate density \citep{Bouwensetal2010,MadauDickinson2014}, stellar mass estimates \citep{Santinietal2015, Momchevaetal2016}. In fact, a lot has been gleaned about galaxy evolution over a period of more than ten billion years from these high quality datasets \citep{Vandenbergh2002,Leitneretal2012,Conselice2014,Pacificietal2023}. Recent JWST observations have pushed the boundary of our knowledge of galaxy formation and evolution with several newly detected galaxies at redshift $z>10$ using its NIRCam and NIRSpec instruments \citep{Eisensteinetal2023,Leethochawalitetal2023,Simonsetal2023,Arrabaletal2023}. 

\begin{figure}
\rotatebox{0}{\includegraphics[width=0.5\textwidth]{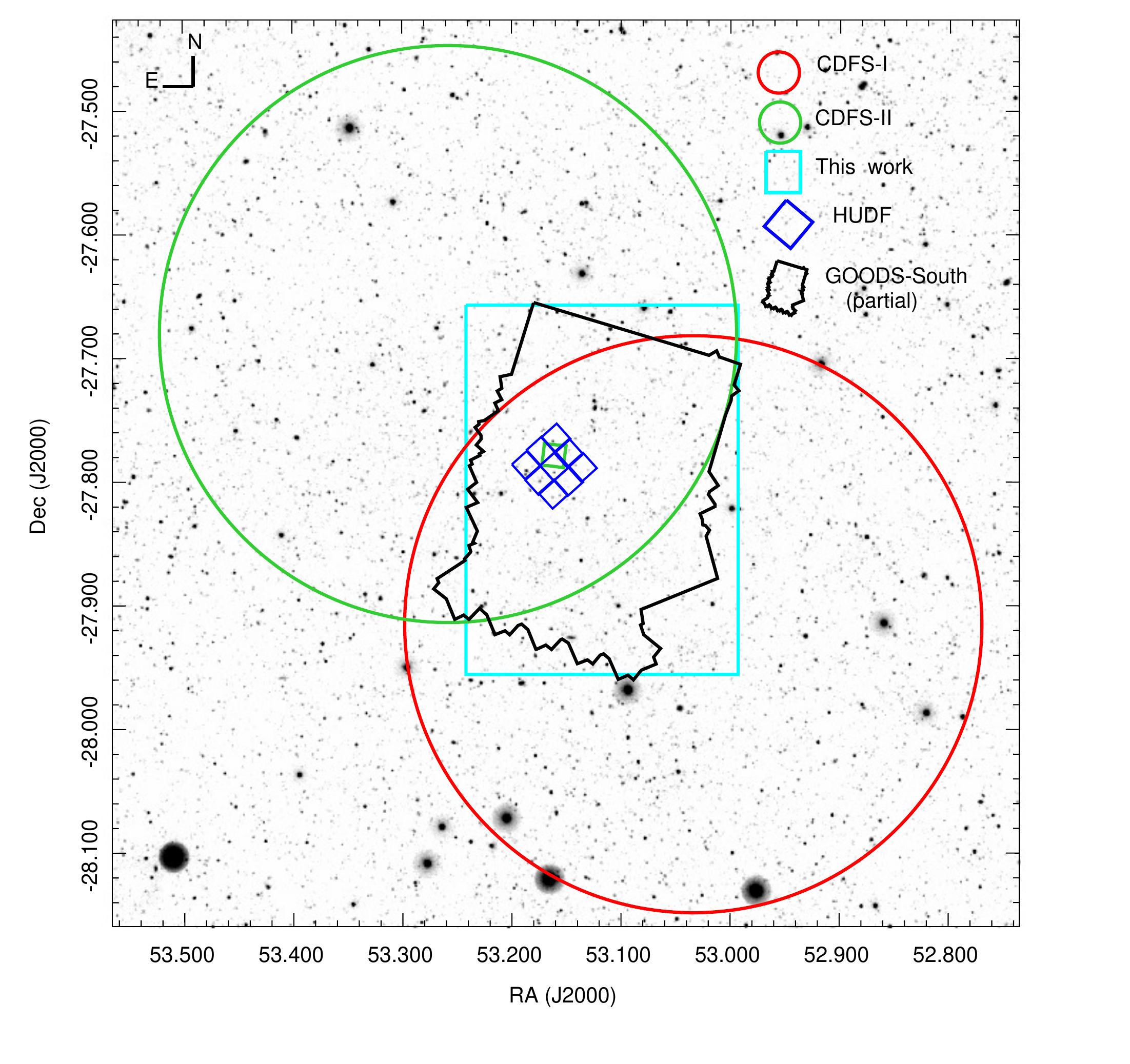}}
\caption{UVIT/AstroSat observational Layout under GT05-240 proposal. CDFS-I (53.03394, -27.91539) and CDFS-II (53.25806, -27.68067) are the two pointing planned for 50 ks on-source observation each in F154W and N242W filters. The blue tiles represent the HUDF \citep{beckwith2006}; each small box has a size of $1' \times 1'$. The green region inside HUDF is known as the Hubble XDF \citep{Illingworthetal2013}. The background image is from GALEX NUV deep field \citep{martin2005}.}
\label{fig:obsplan}
\end{figure}

\begin{figure*}
\centering
\rotatebox{0}{\includegraphics[width=0.8\textwidth]{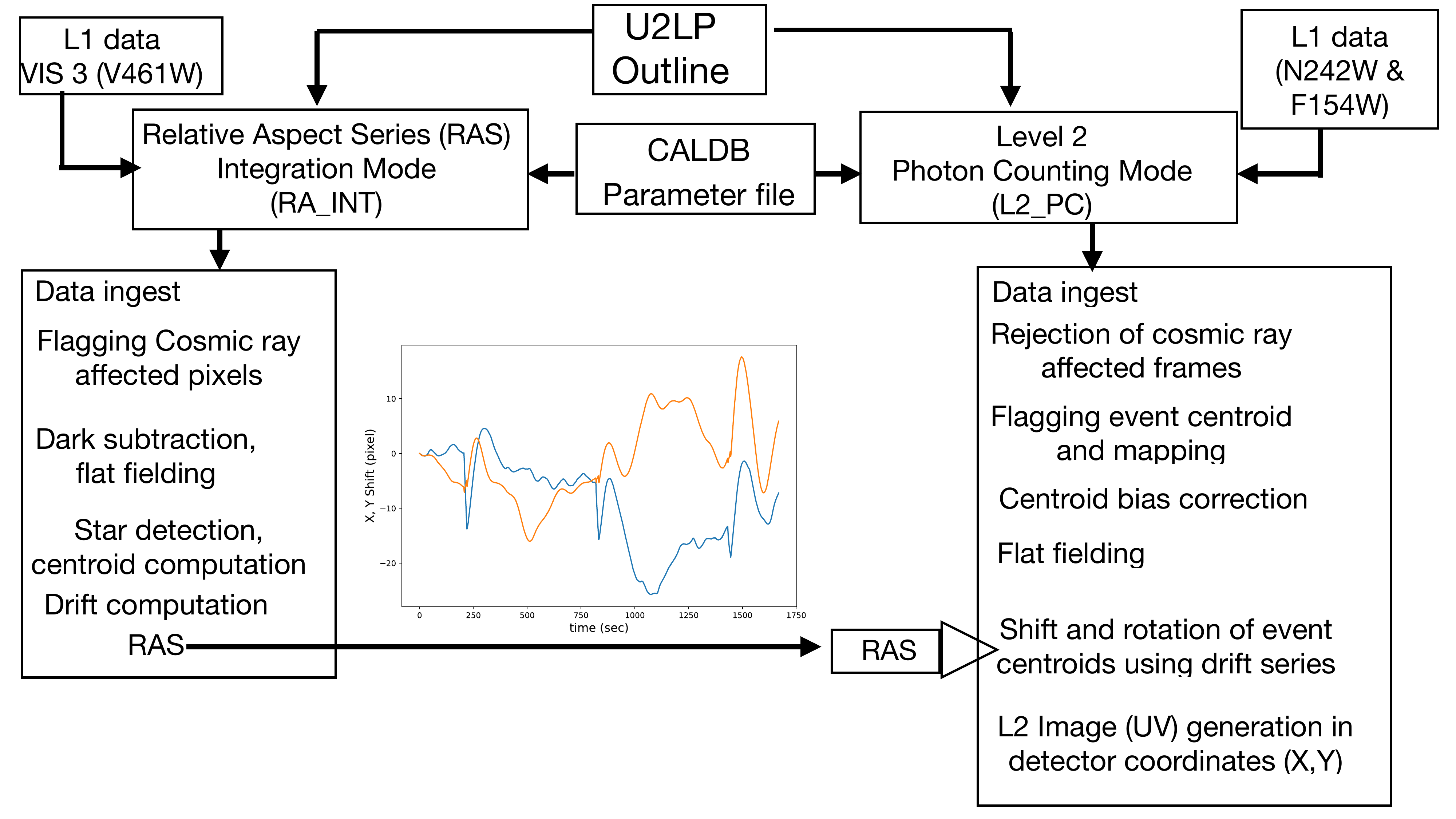}}
\caption{Brief Outline of the U2LP pipeline and key processes involved in creating the orbitwise images in F154W and N242W filters. Inset figure shows the Relative Aspect Series (satellite drift correction) derived from the V461W (VIS3) channel for a single orbit.}
\label{fig:u2lp}
\end{figure*}

On the other hand, deep UV imaging is a great way to directly probe how galaxies grew their stellar mass via in-situ star-formation. But there is a dearth of high-resolution far-UV (FUV; $\lambda \sim 1300 - 1800$\,\AA) and near-UV (NUV; $\lambda \sim 2000 - 3000$\,\AA) deep and wide-field observations covering the entire GOODS-South field at present. Although, the deep observations by GALEX \citep{martin2005} in the FUV and NUV bands have covered the entire GOODS South field, its large point spread function (PSF) with full-width at half-maximum (FWHM) of $\sim 5"$, rendered the faint UV sources (of magnitude beyond its confusion limit $\sim$25 AB mag) unreachable. Over the last decade or so, the HST/WFC3/UVIS channel with its three UVIS filters - F225W; $\lambda \sim 1990 - 3000$\,\AA, F275W; $\lambda \sim 2284 - 3123$\,\AA, F336W; $\lambda \sim 3015 - 3708$\,\AA~ has observed different parts of the GOODS South field with unprecedented sensitivity and resolution e.g., the Early Release Science (ERS, $\sim 47$ arcmin$^2$) field covering the northern side of GOODS South \citep{Windhorstetal2011}; the Ultraviolet Ultra deep field (UVUDF) covering $7.3$ arcmin$^2$ area of HUDF \citealt{teplitz2013, Rafelskietal2015}. More recently, WFC3/UVIS channel has covered $43.4$ arcmin$^2$ area of the CANDELS Deep South producing the Hubble Deep UV Legacy survey (HDUV) in two filters F275W and F336W \citep{Oeschetal2018} and the UVCANDELS in F275W filter covering $\sim 52$ arcmin$^2$ of GOODS South \citep{Wangetal2023}. Most of these deep UV observations are in the wavelength range $2500 - 3500$\AA, probing rest-frame FUV emission only for galaxies beyond $z>0.7$. Nevertheless, these deep UV imaging observations with high spatial resolution and sensitivity have enabled several interesting science goals: the detection of sub-kpc clumps in star-forming galaxies \citep{EE2014,Martinetal2023} in the NUV band at redshift $z\sim 1$ to understand a galaxy's stellar mass build-up; better estimate of the photo-z by including spectral breaks (e.g., Lyman breaks, Balmer breaks) in the UVIS and optical filters \citep{Rafelskietal2015}; constraining the slope of the UV LF beyond $z \sim 1$ \citep{oesch2010, Alavietal2016, Mehtaetal2017}, a search for the Lyman continuum (LyC) leakers \citep{Rutkowskietal2017,Smithetal2018,Wangetal2023}, a direct probe into the cosmic star-formation history which had its peak during $z \sim 1 - 3$ \citep{MadauDickinson2014} and when galaxies made most of their stars. 

However, at relatively lower redshift, say $z < 0.8$, there is a paucity of wide-field, deep, high-resolution UV observation, especially the stellar continuum at rest-frame $1500$\AA~, which is crucial to understand the most recent star-formation activity as it traces young hot stars e.g., O, B or OB association in those galaxies. The imaging in the FUV band (HST/SBC/F150Lp, $\lambda \sim 1420 - 2000$\AA) did not reach adequate depth and sensitivity (due to high dark current in the old MAMA detector, see STScI instrument review documents), and contained very few sources \citep{Sianaetal2007} and did not cover the full GOODS South field. The need of a wide-field imaging observation with reasonable PSF (about an arcsec size or less) and higher sensitivity in the FUV and NUV filters remains. 

\begin{figure*}
\centering
\rotatebox{0}{\includegraphics[width=0.8\textwidth]{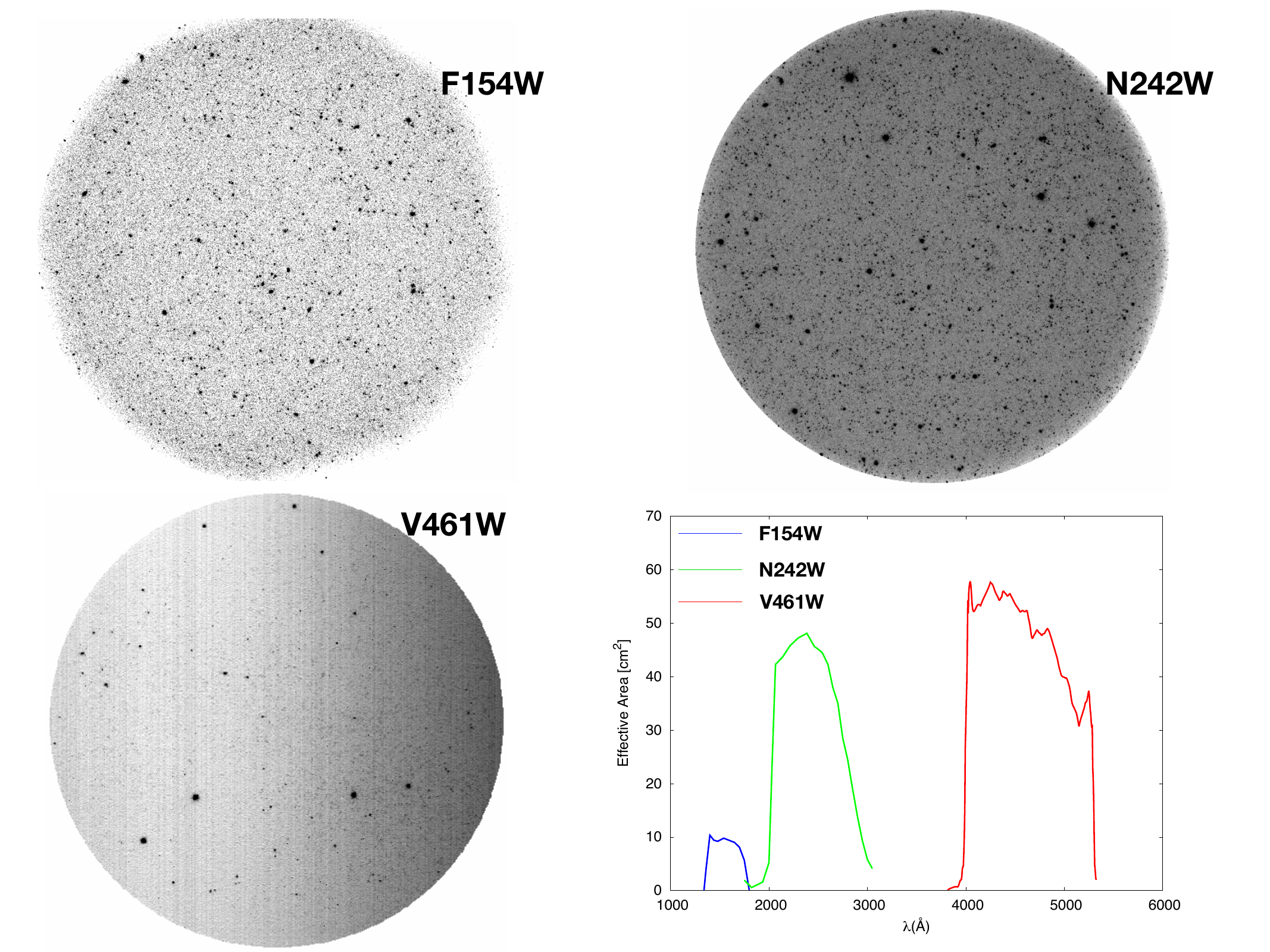}}
\caption{Upper panel: CDFS-II field final image in F154W, N242W filters (see Figure \ref{fig:obsplan} for reference). In the bottom panels are shown the VIS3 (V461W) channel image in a single frame and the Effective area curves as a function of wavelength for the three filters used for observations. The filter effective area is derived by multiplying the filter transmission curve with the quantum efficiency of the detector.}
\label{fig:cdfsII}
\end{figure*}

The Ultraviolet Imaging Telescope \citep[UVIT;][]{Kumaretal2012,Tandonetal2017a} on AstroSat \citep{singh2014} provides an intermediate solution to fill this gap because it has a large circular field of view (28\arcmin ~in diameter) and two co-aligned telescopes capable of performing simultaneous observations in far and near-ultraviolet bands \citep{Tandonetal2020}. The $\sim 1"$ FWHM spatial resolution of UVIT and better sensitivity ($3\sigma$ detection limit $\sim 27.5$ AB mag) provides a unique opportunity to investigate the aforementioned science goals and more, on a larger sample of galaxies. Using UVIT deep observations on GOODS North \citep{Mondal23}, it has been possible to directly constrain the faint-end slope of the UVLF until $z\sim 1$ \citep{Bhattacharyaetal2023} and thereby complementing the UVLF from $z = 2.5 - 0.7$ using WFC3/UVIS filters on HST \citep{oesch2010,Sunetal2023}. This dataset also provided a direct measurement of the UV continuum slopes and its relation to the infrared excess (IRX) from a sample of galaxies from $z \sim 0.4 - 0.8$ \citep{Mondaletal2023, MondalSaha2024}. The deep FUV and NUV wide-field observation of the GOODS South presented in this work has already produced interesting results: \cite{Sahaetal2020} presented an LyC leaker from the HUDF region at $z\sim 1.42$ using the FUV filter probing rest-frame $600$ \AA~ photons and thereby shed some light on the ionizing spectrum of the leaking star-forming (SF) galaxies.  
\cite{Borgohainetal2022} showed extended FUV emission (many of them were clumpy) beyond the optical boundary in 10 blue compact dwarf galaxies at $z \sim 0.1 - 0.3$ in the GOODS South - suggesting clump migration in action to grow these dwarfs. In fact, our deep wide-field UV data on faint SF galaxies can be used to estimate/constrain the Cosmic UV background and help strengthening our understanding of dwarf galaxy formation \citep{Bowyer1991,Voyeretal2011,Bullock2017}. Last but not the least, our deep UV data may be used to have a better estimation of photometric redshift by using the UVIT/FUV band, HST/WFC3/UVIS/F336W and ACS optical filters to probe the Lyman break and Balmer breaks simultaneously in the redshift range $0.07 < z < 0.7$ for the entire GOODS South field \citep{Rafelskietal2015}. Many of these science goals and much more can be achieved by combining galaxies observed in the UV wavelength with optical and infrared bands e.g., the recent GOODS South photometric catalog by \cite{Whittakeretal2019}. However, due to the widely varying point spread function (PSF) and sensitivity, making a homogeneous multi-wavelength catalog using data from different telescopes is a challenging task. In this work, we restrict to a relatively simpler task of providing a comprehensive catalog of galaxies in the far and near-UV bands as well as a catalog of cross-matched galaxies with HST/CANDELS observation \citep{grogin2011,koekemoer2011}. The far and near-UV catalog presented in this work is expected to be a useful value-addition to the astronomy community. 

The rest of the paper is organized as follows: Section~\ref{sec:obs} provides an overview of the planning of the observations with UVIT and data reduction process to final science-ready images. Section~\ref{sec:background} discusses at length the background and noise estimate from the observed multi-orbit data. In section~\ref{sec:PSF}, we detail the PSF analysis, while section~\ref{sec:Sextractor} is on the source extraction, photometry, and catalog preparation. Completeness analysis of the catalog is presented in section~\ref{sec:Completeness}. Section~\ref{sec:imagecomp} displays some of the interesting sources with their optical/IR and UV counterparts and imaging quality. Finally, a summary and conclusions from this work are presented in section~\ref{sec:summary}. We present a thorough analysis of the source extraction in the Appendix. 

A flat $\Lambda$~CDM cosmology with H$_\textrm{0}$ = 70 km$s^{-1}$ Mpc$^{-1}$ , $\Omega_\textrm{m}$ = 0.3, and $\Omega_{\Lambda}$ = 0.7 was adopted throughout the article. All magnitudes quoted in the paper are in the AB system \citep{Oke1983}.

\section{Observation planning and Data processing}
\label{sec:obs}

 The GOODS-South field \citep{dickinson2003} was chosen because of the availability of a large volume of high-quality archival space-based and ground-based data such as HST/ACS BViz imaging \citep{giavalisco2004}; HST/WFC3 UVIS and IR observations; deep VLT/ISAAC JHKs-band imaging \citep{Retzlaffetal2010}; Spitzer/IRAC and MIPS imaging \citep{giavalisco2004,Stefanonetal2021}; Herschel FIR \citep{Elbazetal2011}, ALMA \citep{Franco18} and more recently, JWST near-IR to mid-IR imaging \citep{Gardneretal2006}. In addition, the field is enriched with spectroscopic observations e.g., GOODS/VIMOS \citep{Balestraetal2010}, 3DHST grism survey \citep{Momchevaetal2016} prior to our observations, and more recently, with deep IFU surveys using the Multi-Unit Spectroscopic Explorer (MUSE) on the VLT e.g., MXDF  \citep{Baconetal2017,Herenz2017a,Urrutia2018,Baconetal2022}, VLT/VIMOS \citep{LeFevre19},  VANDELS \citep{Pentericcietal2018, McLure18}.

The FUV and NUV imaging data come from UVIT/AstroSat observations of the Chandra Deep Field South (ObsID: GT05-240, PI: Kanak Saha). The observations were planned with two pointing (called CDFS-I and CDFS-II, see Figure ~\ref{fig:obsplan}) per filter, such that both fields of view (FOV) of size 28\arcmin diameter have an overlapping region containing the GOODS-South field. For each pointing, a total of 50 ks was allocated with the aim that the overlapping region of the FOV would have $\sim 100$ ks of observing time. The UVIT on AstroSat performed these observations simultaneously in far-UV (F154W: $\lambda \sim 1300 - 1800$\,\AA) and near-UV (N242W: $\lambda \sim 2000 - 3000$\,\AA) filters in Photon Counting (PC) mode during Sept. - Oct. 2016 (the early phase of UVIT operation under the Guaranteed Time Observations). 
The orbital period of AstroSat is $\sim 97$ minutes and only about 20 - 30\% of it is available to observe the field during each orbit - corresponding to about $\sim 1100 - 1700$\,sec ($\sim 1400$\,sec on average) for a good orbit. In a given orbit, the available observing time is constrained primarily due to the requirement of being on the dark side of the orbit to avoid scattered light from the Sun and avoidance of the South Atlantic Anomaly. As a result, either the complete dark side of the orbit or a fraction of it is available depending upon the initial settings. During each orbit, FUV and NUV observations were carried out simultaneously in PC mode with frame-rate of 28.7 per second - resulting in about $\sim 30000 - 45000$ frames (in each band) accumulation in a good orbit. Our imaging observations were carried out as a sequence of imaging sessions with a window size of $512 \times 512$~pixels and raw data i.e., science data in each filter and Auxiliary data from the satellite were stored orbit by orbit.  

In addition to the FUV/NUV observations, UVIT also carried out observations in the VISUAL band using VIS3 channel (V461W: 3878 - 5325 \AA) in the integration mode (IM) with a frame rate of 1 frame per second during the same period. The total exposure time for the V461W filter was 50 ks on-source for each pointing. Apart from the UV observations, AstroSat also observed both the proposed fields with the Soft X-ray Telescope (SXT, 0.3 - 8 keV) in photon counting mode \citep{KPSinghetal2017}; Cadmium-Zinc-Telluride Imager (CZTI, 10 - 150 keV) in normal mode \citep{Bhaleraoetal2017} and the Large Area X-ray Proportional Counter (LAXPC, 3 - 80 keV) in default event analysis mode \citep{Yadavetal2016} over the full duration of the observation stated above. These observations will be presented in a separate future paper.   

\begin{figure}
\includegraphics[width=0.5\textwidth]{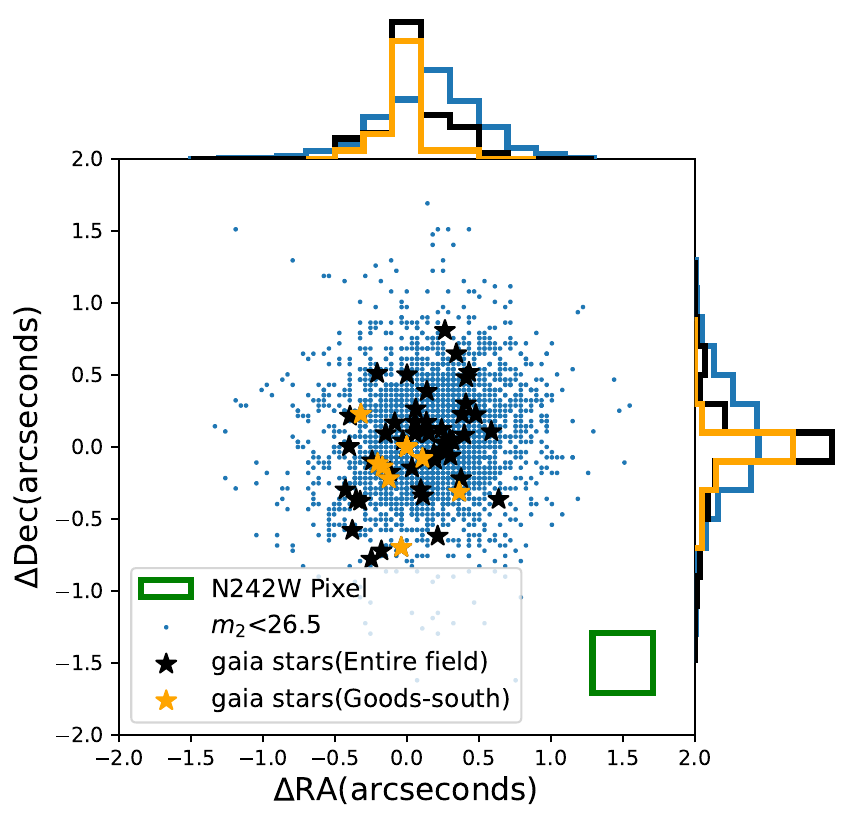}
\caption{Difference between the coordinate of a star in the GAIA EDR3 catalog and its location on the N242W image. The orange solid histograms represent the difference in RA and Dec for the stars that belong to the GOODS South region only while black histograms for the entire field (CDFS-I and CDFS-II). The blue dots represent the same but for randomly selected bright (N242W$\_ $m2  $<26.5 )$ objects with their unique counterparts marked as CANDELS\_Flag=1 in {\it AUDFs-N242W-Cat1-v1}. The RMS differerence for the GOODS South region is 0.13\arcsec and 0.18\arcsec along RA and Dec respectively. While for the entire region they are 0.38\arcsec and 0.34\arcsec respectively. The green square represents one UVIT pixel. }
\label{fig:diff_coord}
\end{figure}

\begin{figure*}
\rotatebox{0}{\includegraphics[width=0.9\textwidth]{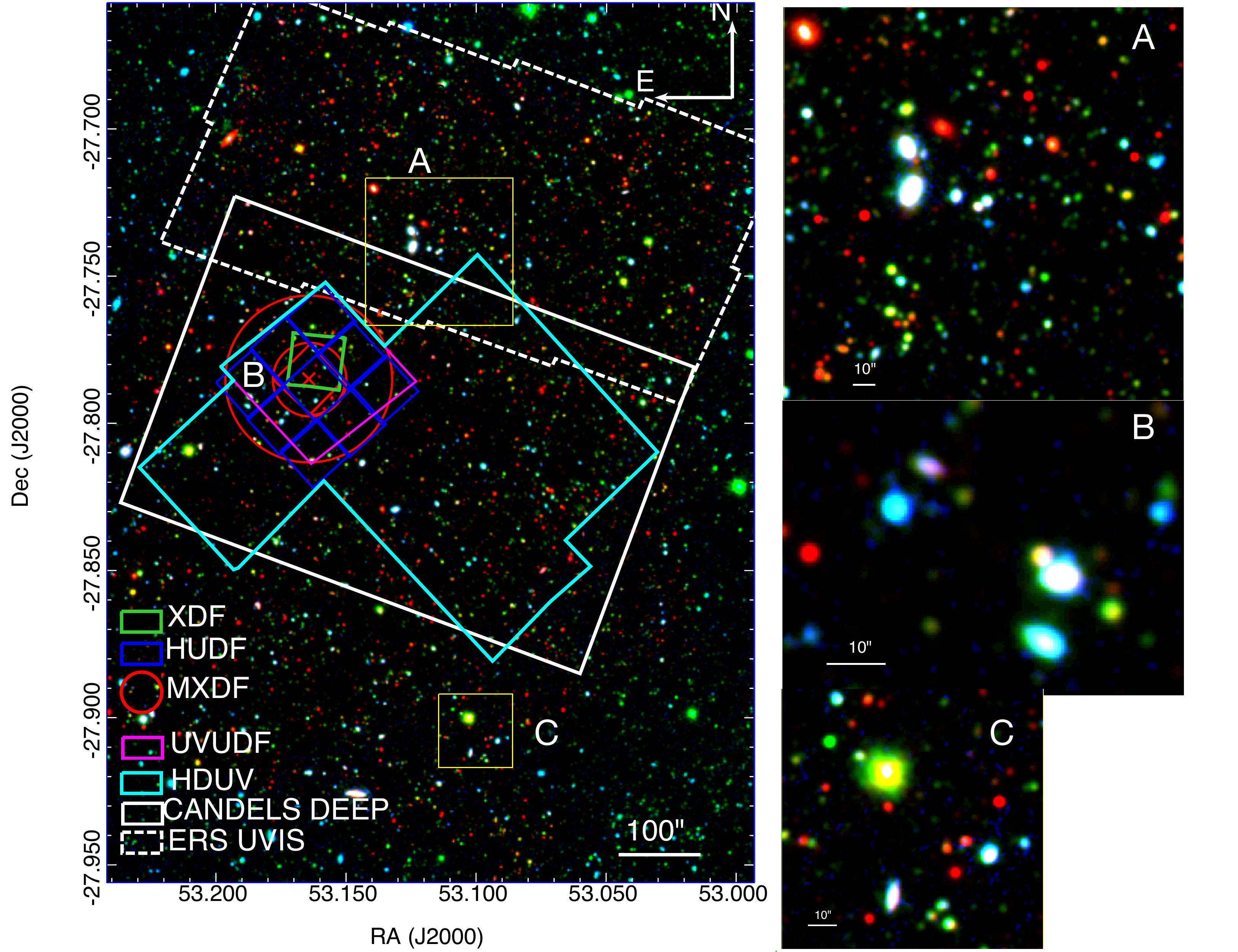}}
\caption{False color image of the AUDF South. Red refers to the HST/ACS/F606W image; green the N242W and blue the F154W filter of UVIT on AstroSat. HST and N242W images are convolved with F154W PSF and have the same pixel scale as in the F154W band. The zoom-in portion (A, B, and C) of the image is shown to highlight the imaging quality and astrometric accuracy of the two UVIT filters. The region B refers to one of the blue squares from HUDF.}
\label{fig:rgb}
\end{figure*}

\subsection{Orbit by orbit data reduction}
We obtained the complete raw dataset as Level 1 (L1) FITS files from the AstroSat archive  \footnote{https://astrobrowse.issdc.gov.in/astro-archive/archive/Home.jsp} \citep{Balamuruganetal2021} during January, 2017 and processed the L1 data to create Level 2 (L2) science-ready images using an earlier version of the official UVIT L2 pipeline (hereafter UL2P, \cite{Ghoshetal2022}).  The L1 dataset for every orbit contains the science data from each filter (i.e., F154W, N242W, VIS3) and the Auxiliary data which provides critical information such as the time calibration table, orbital position, attitude information, housekeeping information of the satellite during the entire period of observation. 

We briefly outline the processes (see Figure~\ref{fig:u2lp}) involved in the conversion of raw L1 data to L2 as it applies to our observations which have VIS, NUV and FUV band data; for details, the readers are referred to \cite{Ghoshetal2022}. U2LP consists of two primary chains of operations - derivation of the Relative Aspect Series in integration mode (RA-INT) and the Level 2 image generation in photon counting mode (L2-PC). During the RA-INT, the pipeline works on the VIS3 (V461W) data along with the CALDB (instrument calibration database) and a user-supplied parameter file to generate satellite drift correction. It flags the cosmic ray-affected pixels by identifying if a pixel value is higher than a selectable threshold (more like removing hot pixels), then performs dark subtraction and flat fielding. In the next step, it detects stars in each image and corrects their centroid position to account for the instrument effects. Finally, the RA-INT computes the shift in the star centroids in detector coordinates (X, Y) as well as in the spacecraft coordinates as a function of time during the orbit (see the inset in Figure ~\ref{fig:u2lp}) and this is stored as a drift series (i.e., RA series for each orbit) to be fed to the second chain called L2-PC. During the second chain (generation of UV sky image), the pipeline invokes the L1 data (both filters F154W and N242W) along with CALDB and the parameter file (as in RA-INT). After the data-ingest process (i.e., collecting the list of centroids of photon events; note the centroids are calculated to 1/32 of a pixel of the original $512\times512$ CMOS imager, see \cite{Tandonetal2017a}), it identifies the cosmic ray (based on a user-provided threshold mentioned in the parameter file) affected frames and rejects them all. 

Often, cosmic rays appear as a characteristic shower which affects a large portion of the image. For a given observation (consisting of multiple orbits), such cosmic ray-hit frames may lead to $\sim 10$\% data loss (as the pipeline completely removes the affected frames), effectively reducing the final exposure time. In other words, those frames were excluded in the final science-ready images and the subsequent calculation of photometry. Additionally, in our data reduction, there was also data loss due to the mismatch of the time-stamp on V461W (VIS3 channel) and N242W or F154W filters. Besides the identification of cosmic ray-affected frames, the pipeline flags bad pixels, multiple photon events and applies flat fielding. Once this step is completed, the pipeline computes photon event weight and applies correction to photon centroids due to various instrument systematic. This includes pixel padding where photon centroid coordinates are modified from (1 - 512) pixel to (1 - 600) pixel to accommodate spacecraft drift; pixel sub-division of centroids in which all photon centroid coordinates are transformed from (600, 600) to (4800, 4800); centroid bias correction; detector distortion \citep{Girishetal2017}; optical distortion \citep[see][for details]{Ghoshetal2022}.  
Finally, the pipeline applies the drift series (derived from RA-INT) to compute the shift and rotation of the photon event centroid and generate L2 images in the detector coordinate as well as the exposure map for each orbit. In essence, the L2 Pipeline transforms the photon event coordinates to generate the images in an idealized detector X/Y coordinate system, along with data on the exposure for each pixel of the image (note that due to the drift the pixels near the boundary have less net exposure as compared to the centre.) 

At this stage, we extract all the orbit-wise images in each filter and their exposure maps in the detector coordinate system. These images and exposure maps are then combined, applying shift and rotation with respect to orbit 1 image as the reference image. To estimate the shift and rotation, between two consecutive images (e.g., between the reference image and orbit 2 image), we first identify a set of point-like sources spanning over the full FOV in the reference image and their centroid position in the detector coordinate system (X1, Y1). The same set of sources are being identified (visually by pattern matching) in orbit 2 image and their centroid position (X2, Y2) are computed. This process is continued for all the individual orbit images for the full duration of the observation keeping orbit 1 image as the reference image.

Often, it is not possible to identify stars in a single orbit FUV image due to the low photon counts and not having enough SNR. Hence, other point-like sources are utilized to create initial alignment of the images from different orbits as explained before. These point-like sources are common between the corresponding GALEX CDFS deep field and our target fields (CDFS-I and CDFS-II, see Figure ~\ref{fig:obsplan}). According to the GALEX PSF encircled energy (EE) curve \citep{Morrisseyetal2007}, a point source contains about 70 -80\% of the total light within an aperture of $\sim 5$\arcsec (FWHM of GALEX PSF). For all practical purposes, we utilized GALEX PSF EE to define and select a point-like source. We selected 8 sources from CDFS-II of which one is an actual star and 7 from CDFS-I field of which 3 are actual stars; rest are likely to be galaxies which appear point-like in GALEX. The centroid positions are calculated by placing a suitable box around each source and calculating their centroid of intensity. We use the Newton-Gauss iterative method to estimate the reciprocal translation and rotation between the two images (this part is done using an IDL code as part of the NASA IDL Astronomy library\footnote{https://idlastro.gsfc.nasa.gov}). This process is applied to the rest of the orbit images; the exposure map for the corresponding orbit is simply subjected to the same shift and rotation. The resulting intensity map ($I_\textrm{k}$) and exposure map ($T_\textrm{k}$) are matrix multiplied and summed over all orbits (N$_\textrm{orbit}$) as shown in the following equation:  

\begin{equation}
\tilde{I}_\textrm{m}=\frac{\sum_\textrm{k=1}^{\textrm{N}_\textrm{orbit}}{{\Delta t}\times T_\textrm{k} I_\textrm{k}} } {\sum_\textrm{k=1}^{\textrm{N}_\textrm{orbit}}{{\Delta t}\times T_\textrm{k}}},
\end{equation}

\noindent where the units of $I_\textrm{k}$ and $T_\textrm{k}$ are counts per second (cps) and number of frames. $\Delta t=1/28.7 = 34.8$~second denotes the time-interval between two consecutive frames and $\tilde{I}_\textrm{m}$ is the final image in cps. We followed the above procedure to obtain images of CDFS-I and CDFS-II fields (see Figure ~\ref{fig:obsplan}). Often all the orbit combined image has a size slightly larger than $14\arcmin$ radius. Since image artifacts are expected to concentrate near the periphery of the FOV \citep{Morrisseyetal2007}, we restrict only to $14\arcmin$ radius from the center of the FOV. Figure ~\ref{fig:cdfsII} displays the combined (all orbit) images of CDFS-II (one of the two proposed fields) in F154W and N242W filters. We also display the single-frame VIS3 (V461W) image, for illustration purpose only. The filter effective area (which is derived by multiplying the filter transmission curve and its quantum efficiency) curves for all three filters used in this work are also shown in the same figure. Note the mean effective area in F154W and N242W filters are 10 cm$^2$ and 40 cm$^2$ respectively \citep[see][]{Tandonetal2017a}.   

\subsection{Astrometric alignment}
\begin{figure}
\begin{center}
\rotatebox{0}{\includegraphics[width=0.5\textwidth]{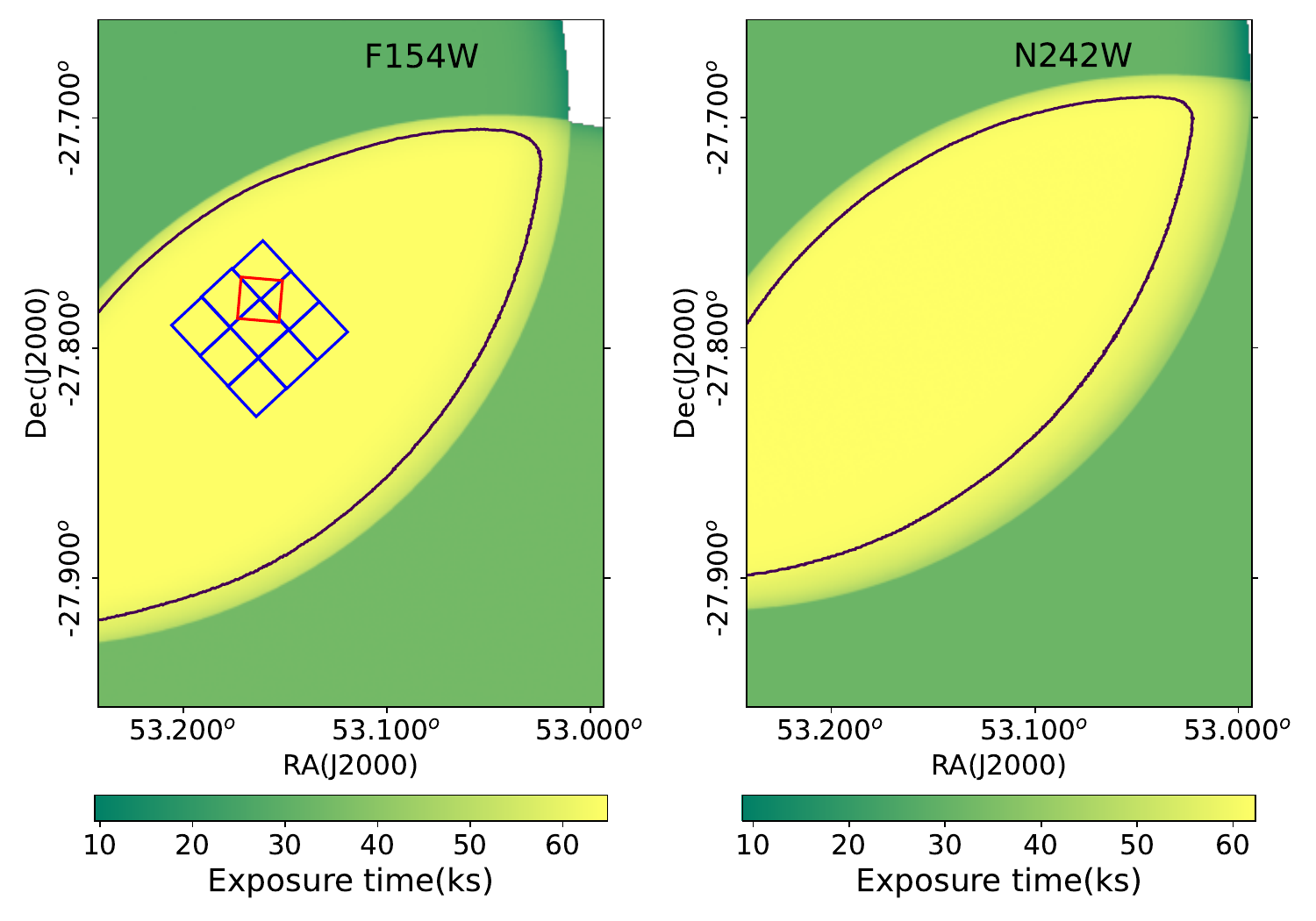}}
\caption{Combined exposure maps: in the overlapping region, the exposure time is the highest. On either panel, the contour is drawn at 60 ks ($\sim 43$ orbits on average). At the edge of the FOV, the exposure time drops drastically. Outside the overlapping region and within 1$\arcmin$ from the edge of the FOV, local exposure time is greater than 30 ks in either case. The footprint of the UDF mosaic is shown for reference only.}
\label{fig:expmap}
\end{center}
\end{figure}

Both CDFS-I and CDFS-II (see Figure~\ref{fig:obsplan}) were subjected to astrometric alignment which was performed in two steps. First, we used GALEX deep field data (in FUV and NUV) of the GOODS-South region and identified a set of point sources (explained above) for which we have x pixel/y pixel from {UVIT} image and their corresponding RA/Dec from GALEX field. We use an IDL program that takes these inputs and performs a TANGENT-Plane astrometric plate solution similar to ccmap task of IRAF\footnote{https://iraf-community.github.io/}. We then combine CDFS-I and CDFS-II fields using SWARP\footnote{https://www.astromatic.net/software/swarp/} to create a combined image in both filters. Apart from star-tracking, we do not use the V461W image for any photometric measurement. In the second step, we identify 93 GAIA \citep{GaiaEDR32021} stars on the combined image and match their position to its NUV counterpart in the N242W filter to redo the astrometric calculation for the final correction. In the case of the F154W filter image, we could identify 59 GAIA stars and followed the same procedure (as it was in the first step) to perform the astrometric correction. All these GAIA-UVIT matched stars lie in the full FOV represented by CDFS-I and CDFS-II (see Figure ~\ref{fig:obsplan}). 
The final astrometric accuracy in the N242W and F154W filter images has an RMS$\sim 0.2"$ (note, a pixel $\simeq 0.4"$ in UVIT sub-pixel images). A rectangular patch containing most of the GOODS-South region (indicated by cyan color in Figure~\ref{fig:obsplan}) is cut out and presented in the rest of the paper. The patch does not cover (Eastern side) the whole overlapping region as there is no HST data. The GAIA star positions and their counterparts in the UVIT image are shown in Figure~\ref{fig:diff_coord}, highlighting the astrometric accuracy; note that there are only two GAIA matched stars in the GOODS-South region (rectangular patch, marked in Figure~\ref{fig:obsplan}) in the F154W filter. The rest of the F154W filter matched stars are outside this rectangular patch. The final science-ready image (false color RGB: AstroSat/F154W, N242W and HST/F606W) of the AstroSat UV deep Field (AUDF) presented in this work is shown in Figure ~\ref{fig:rgb}; while the exposure maps for the two filters are shown in Figure ~\ref{fig:expmap}. Some offsets are visible in the zoom-in RGB images displayed in Figure ~\ref{fig:rgb} (A,B,C), although these offsets are within the astrometric error (see also text at the end of Section~\ref{sec:Sextractor}). The overlapping region has a higher exposure time (by a factor of $\sim 2$) than individual stacked field. The exposure times in the overlapping region are $\sim 64$~ks and $\sim 62$~ks in the F154W and N242W filers respectively. The overlapping and the non-overlapping regions in each filter image are termed as deep and shallow regions respectively.
The photometric calibration is performed with a white dwarf star Hz4; the updated photometric zeropoints \citep{Tandonetal2020} are 17.77 and 19.76 for F154W and N242W respectively. 

\section{Background measurement}
\label{sec:background}

The measured background from the science-ready image has contributions from the instrumental noise which is primarily detector dark counts in case of UVIT and the actual background from the sky. A thorough measurement of the sky background (global as well as local) in the UV is necessary for measuring accurate source fluxes, especially for the objects in the fainter regime. The clipping of pixels to reduce source contamination is a traditional technique for background estimation \citep{Almozninoetal1993}, with a variant of this being used in the source detection software, Source Extractor (SExtractor) \citep{bertin1996}. From several tests varying the SExtractor key detection parameters, we found that it overestimates the true background in a source-crowded region. Moreover, SExtractor appears to fail the background estimation \citep[see][in the case of GALEX background]{Morrisseyetal2007} in low photon count images like the far-UV filter F154W, where the mean full exposure photon count of the background ($\lambda \times  t_\textrm{exp}$) is $<1$~per pixel, giving background values absurdly small (in order of $10^{-9}$ cps/pixel or so). 

\begin{figure*}
\centering
\includegraphics[width=0.8\textwidth]{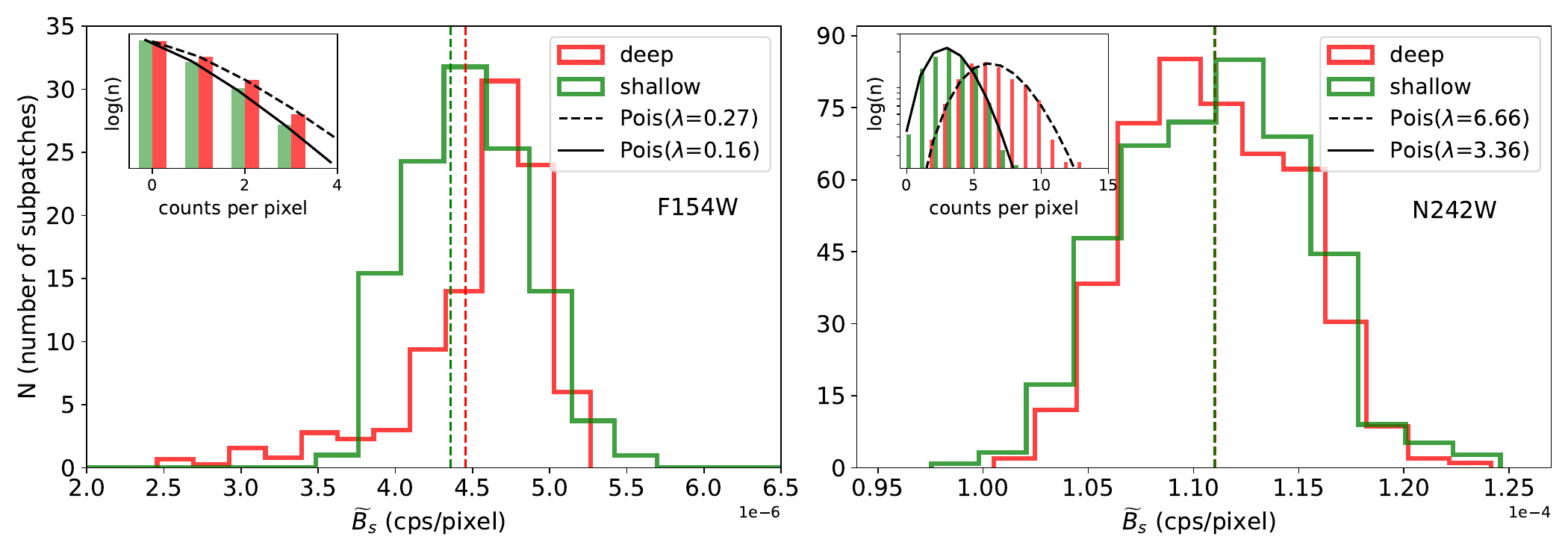}
\includegraphics[width=0.8\textwidth]{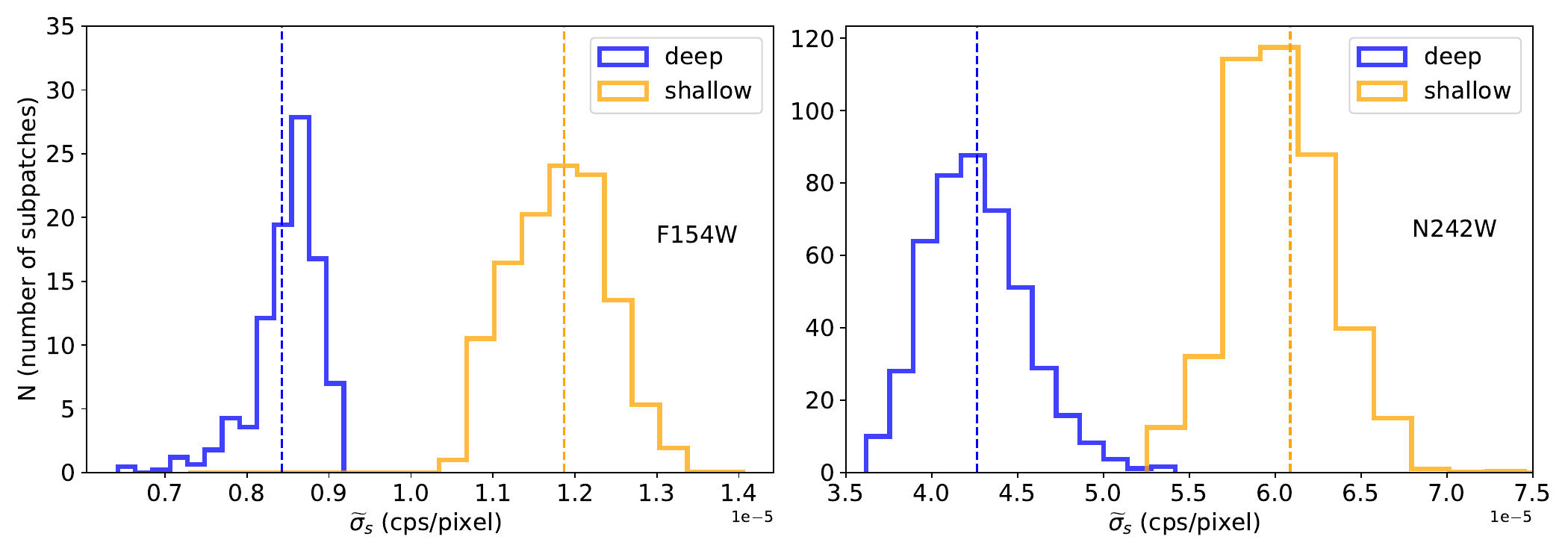}
\caption{Upper panels: histograms of mean background in F154W and N242W. Bottom panels: standard deviation in F154W, and N242W filters. The dashed lines represent the mean values provided in table \ref{tab:back}. The sample background pixel histograms of a randomly selected sub-patch from each deep and shallow part of N242W and F154W with full integration time ($t_\textrm{exp}$) are shown as inset plots in the top left corners of upper panels; their distribution clearly suggesting their Poisson nature.}
\label{fig:back_hist}
\end{figure*}

In such low-photon count images, a semi-random technique of placing boxes in empty regions (in apparent source-free regions i.e., away from SExtractor segmented area) is usually adopted \citep{Rafelskietal2015}. We use a variant of the FellWalker \citep{Berry2015} based (intensity) minima finding method (Pushpak Pandey et al. in prep), to find such relatively source-free regions and place boxes automatically to calculate the background. The whole rectangular patch of image shown in Figure\ref{fig:rgb} was divided into sub-patches of size $74\times 74$~pixels ($\sim 30.8\arcsec\times 30.8\arcsec$) for N242W  and $144\times 144$ pixels($\sim 60\arcsec\times 60\arcsec$) for F154W. This results in $\approx$ 850 and $\approx$220 sub-patches for the N242W and F154W filters respectively.
In each sub-patch, we find local minima and place boxes of size $5\times5$ 
 $(6\times6)$~pixels (larger than PSF FWHM) for the deep (shallow) region. Typically, we are able to place $\sim 7 - 12$ such boxes for N242W and $\sim 40 - 50$ for F154W depending on the crowding or presence of bright, extended sources. For example, if a sub-patch contains a source whose size is similar to the box size, approximately $7 (20)$ nearest boxes are collected from the surrounding region of the N242W (F154W) image. In this way, a total of approximately $8000$ such boxes are placed all over the rectangular area (see the cyan region in Figure~\ref{fig:obsplan}) for the background estimation in the N24W and F154W bands. In each sub-patch, the distribution of counts per pixel (from either filter image) collected from the minima boxes follow a Poisson distribution: 

\begin{equation}
    P_\textrm{s}(n;\widetilde{B_\textrm{s}}) = \frac{e^{-\widetilde{B_\textrm{s}}} {\widetilde{B_\textrm{s}}}^{n}}{n!},
    \label{eq:poisson}
\end{equation}

\noindent $n$ takes integer values representing the number of detected photons and $\widetilde{B_\textrm{s}} >0$ is the mean of this distribution. By definition, the standard deviation of the photon count distribution is $\sqrt{\widetilde{B_\textrm{s}}}$. In units of counts/pixel/sec, the mean background and the standard deviation of the sub-patch is related to the sub-patch exposure time simply by the following relations:

\begin{equation}
\widetilde{B_\textrm{s}} = \frac{\widetilde{B_\textrm{s}}}{{\widetilde{t}}_\textrm{exp}} \: \: \: and \: \:
    \widetilde{\sigma_\textrm{s}} =\frac{\sqrt{\widetilde{B_\textrm{s}}}}{{\widetilde{t}}_\textrm{exp}}
    \label{eq:sigma}
\end{equation}

\begin{figure*}
\centering
\rotatebox{0}{\includegraphics[width=0.8\textwidth]{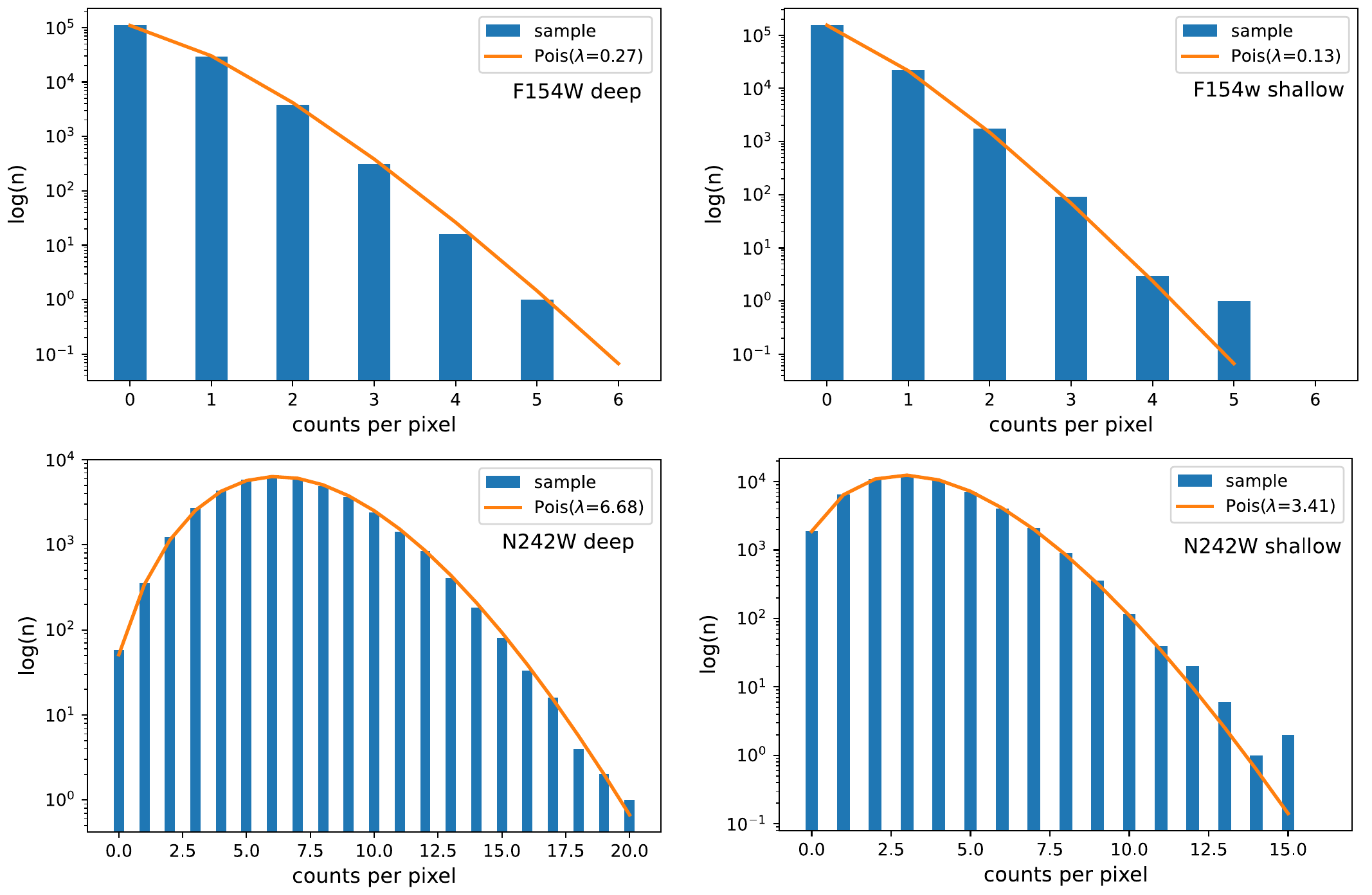}}
\caption{All sampled background pixels from deep and shallow regions of N242W and F154W with full integration time, plotted with the Poisson distribution having $\lambda$ = $\widetilde{B}\times \widetilde{t}_\textrm{exp}$ from table \ref{tab:back}.}
\label{fig:sample_pixels}
\end{figure*}

\noindent The upper panels of Figure~\ref{fig:back_hist} present a set of histograms of the $\widetilde{B_\textrm{s}}$ values calculated from all the sub-patches belonging to the deep and shallow region in each filter. The spread of the histograms suggests spatial variation in the background and the noise over the full field. This was cross-checked visually; see a discussion on the 2D background map, later in this section. The mean values ($95\%$ confidence limit) of these sub-patch histograms are presented in Table~\ref{tab:back}. For clarity, the count distribution for a random sub-patch from the deep and shallow region is displayed as inset figures in the upper panels of Figure~\ref{fig:back_hist}. A Poisson distribution function is fitted to these count distribution and their respective $\lambda =\widetilde{B_\textrm{s}}\times \widetilde{t}_\textrm{exp}$ are shown as legends. It is worth reminding the reader that the exposure time in the deep region is roughly a factor of 2 higher than the shallow region (somewhat reflected in the $\widetilde{B}_\textrm{s}$ values of $0.27$ and $0.16$ in the legends for the deep and shallow of F154W image respectively. For the N242W, they are $6.66$ and $3.36$). 
The histogram distribution of $\widetilde{\sigma}_\textrm{s}$ are shown in the bottom panels of Figure~\ref{fig:back_hist}. The mean $\widetilde{\sigma_\textrm{s}}$ ($95\%$ confidence limit) is displayed in Table~\ref{tab:back}. It is a trivial corollary but at the same time satisfactory to check that $\widetilde{\sigma_s}$ in the deep region is lowered by a factor of $\sqrt{\widetilde{t}_\textrm{exp,deep}/\widetilde{t}_\textrm{exp,shallow}} \simeq \sqrt{2}$ confirming the Posisson nature of the background noise. This holds true for both filter images, i.e., in the F154W and N242W filters. 
\par
In Figure~\ref{fig:sample_pixels}, we show histograms of background pixel counts collected from all the minima boxes in the deep and shallow regions. Each of these histograms is fitted separately with a Poisson distribution with $\lambda = \widetilde{B_\textrm{s}}\times \widetilde{t}_\textrm{exp}$ for cross-checking our calculation. The fitted $\lambda$ values for each filter in the deep and shallow region are given in the legends of Figure~\ref{fig:sample_pixels}) and are used to estimate the global $\widetilde{B_\textrm{s}}$ and $\widetilde{\sigma_\textrm{s}}$. For example, in the N242W deep region, we have $\lambda = 6.68$ and the mean $\widetilde{t}_\textrm{exp} = 60168$~sec which implies $\widetilde{B_\textrm{s}}= 1.11 \times 10^{-4}$ and $\widetilde{\sigma_\textrm{s}}=4.29\times 10^{-5}$ cps/pixel and similarly for the other region and filter. Our global estimate of mean and standard deviation thus closely matches with the estimation from the sub-patch histograms (see Table~\ref{tab:back}).    

Additionally, in order to understand whether there is pixel-to-pixel correlated noise, we inspect the distribution of pixel values. In the F154W band, there are a large number of pixels with `0' and `1' counts as can be seen from the sub-patch pixel distribution (Figure~\ref{fig:back_hist}). The same is true for the N242W filter. This implies no mixing of pixels in our final science-ready images and hence null correlated noise. Further, we estimate $\widetilde{\sigma_\textrm{s}}$ using the same sub-patch method (see Figure~\ref{fig:back_hist}) but on a set of two binned images: $2\times2$ and $3\times3$ pixel binning. For N242W deep region, we obtain $\widetilde{\sigma_\textrm{s}}=8.99\times 10^{-5}$ and $1.29\times 10^{-4}$ cps/binned pixel for $2\times2$ and $3\times3$ binned image respectively. After dividing these $\widetilde{\sigma_\textrm{s}}$ values from the binned images with $\sqrt{4}$ and $\sqrt{9}$, we get $4.49\times 10^{-5}$ and $4.29\times 10^{-5}$ cps/un-binned pixel - close to the numbers presented in Table.~\ref{tab:back}. In other words, these experiments suggest that the increase ($\sim 5\%$ for $2\times2$ and $\sim 0.7\%$ for $3\times3$ binning) in the noise estimate is within the measured uncertainty.    
\par
\begin{figure}
\rotatebox{0}{\includegraphics[width=0.5\textwidth]{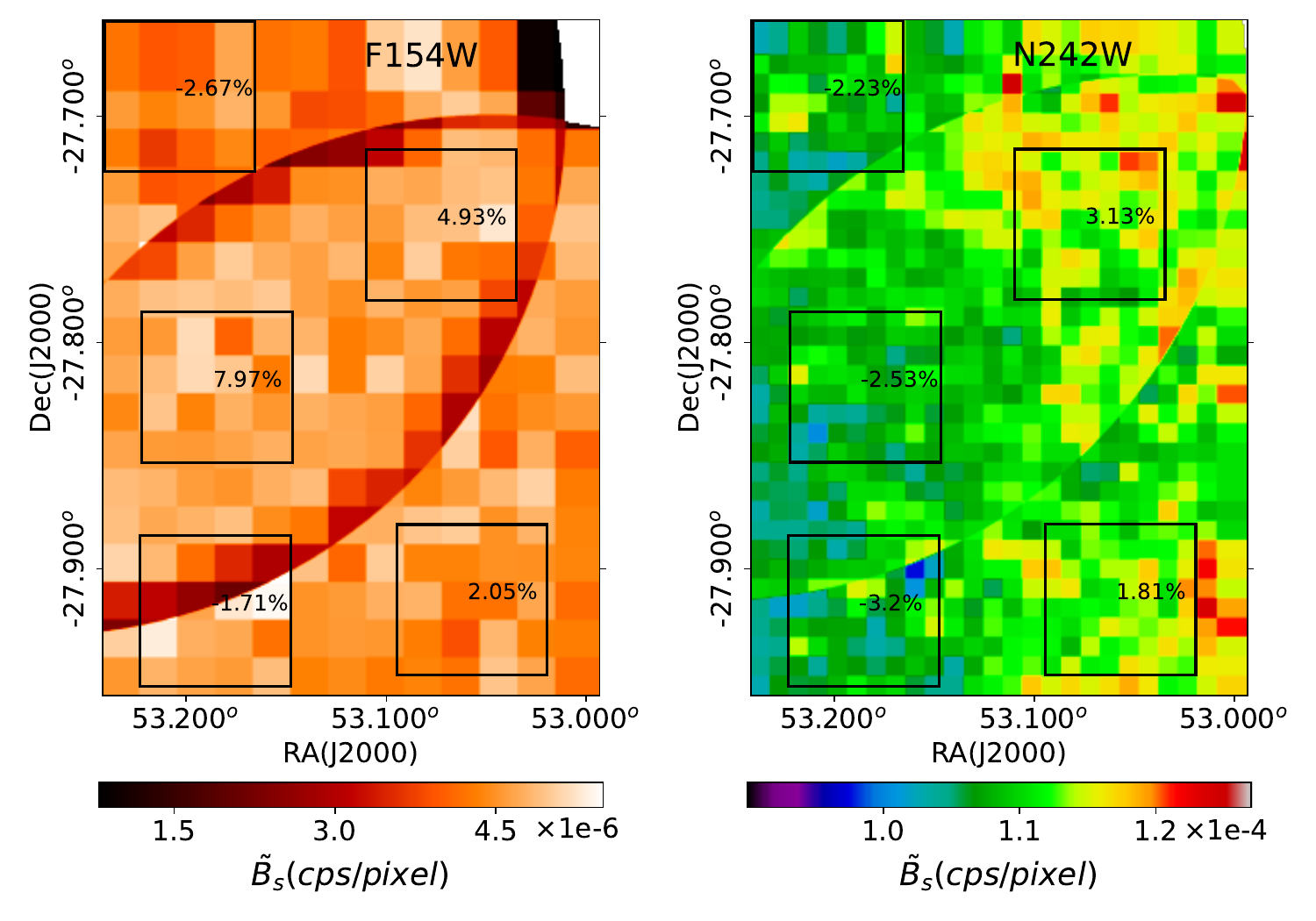}}
\caption{The background map for the F154W (left) and N242W (right). The randomly placed black boxes are of size $4'\times4'$; numbers within represent their percentage deviation from the mean background. Each sub-patch (which resembles a pseudo-pixel in the map) has the size $60\arcsec\times60\arcsec$ and $30.8\arcsec\times 30.8\arcsec$ for F154W and N242W image respectively.}
\label{fig:backmap}
\end{figure}

 
Figure~\ref{fig:backmap} shows the 2D map of the background in F154W and N242W filters. The 2D background map is estimated from the sub-patch mean (see Eq.\ref{eq:sigma}) and this same mean has also been used to estimate the background histogram (see Figure~\ref{fig:back_hist}). In other words, within each sub-patch (e.g., $30.8\arcsec \times 30.8\arcsec$ for N242W image), the background has been assigned with the same mean and standard deviation. We use the standard deviation $\widetilde{\sigma_\textrm{s}}$ values from each sub-patch to create a background noise or root mean square (rms) map.

Overall, there is no apparent gradient in the F154W background but there are fluctuations as high as $\sim 8$\% above the global mean. Fluctuations are also present in the N242W filter but they are about a few percent (max $\sim 3$\%) over the global mean. Our background fluctuations nearly match with background estimate done on the AstroSat UV Deep Field north (AUDFn) by \cite{Mondal23} over the entire field of view. However, for our N242W image, there seems to be a local gradient in the background that roughly correlates with enhanced local source density. This is also true when one runs SExtractor on an optical/IR image - the SExtractor-generated background is higher where there is higher source density. The calculation of local background in a crowded region of the sky is tricky due to unavoidable contamination from nearby sources. We have examined this aspect on a simulated image (with 2000 point sources in 100\arcsec square box with known background). In the crowded region, the overestimation of the background is $<4\%$ (details will be presented by Pushpak Pandey et al. in prep). For a 26 AB mag object, the overestimation of the background (measured within an aperture of radius 1.2\arcsec) would cause an error of less than 0.01 mag.
     

\begin{table*}[]
\caption{Global mean and standard deviation of the background sky as measured in the two UVIT images. The $3\sigma$ detection limits (lim$_{3\sigma}$) are measured within apertures of radius 1.6" and 1.2" for the F154W and N242W bands, respectively. We have used the mean rms value from the third column for its estimation. The 90\% and 50\% detection limits are measured with artificial star tests (see Section~\ref{sec:Completeness}).}
\label{tab:back}
\hspace{-2.5cm}\resizebox{2.5\columnwidth}{!}{%
\Huge
\begin{tabular}{lllllllllll}
\hline
{Telescope}   & {Filter} & Effective  & N$_\textrm{orbit}$ & {$t_\textrm{exp}$} & {$\widetilde{B}$ ($95\%$ Conf)} & {$\widetilde{\sigma}$($95\%$ Conf)}  & {lim$_{3\sigma}$} & {lim$_{90\%}$} & {lim$_{50\%}$}             \\

&        & wavelength  &   & mean (max)         &                     &  &  &  &   \\

                          &        &   (\AA)   &   & \hspace{0.1cm} (second)         & \hspace{0.4cm} { (cps/pixel)}                     & \hspace{0.4cm} {  (cps/pixel)} & {(AB mag)} & {(AB mag)} & {(AB mag)}                            \\ \hline
\multicolumn{1}{c}{AstroSat/UVIT} & N242W  & 2418  & & & & & & &         \\

 & Shallow  & & 22 & \hspace{0.4cm}$30722\ (31733)$        & 1.1E-4 (4.1E-7)         & 6.1E-5 (3.2E-7)      & \hspace{0.55cm}27.34 & \hspace{0.4cm}26.25  & \hspace{0.4cm}27.3         \\

                          & Deep   &   & 44 & \hspace{0.4cm}$60168\ (62345)$        & 1.1E-4 (3.6E-7)        & 4.2E-5 (2.7E-7)    & \hspace{0.55cm}27.73 & \hspace{0.4cm}26.82  & \hspace{0.4cm}27.63           \\ 
                          \hline
AstroSat/UVIT                     & F154W  & 1541  &  & & & & & &             \\

 & Shallow  &  &  25 & \hspace{0.4cm}$31514\ (34791)$        & 4.3E-6 (6.9E-8)        & 1.2E-5 (1.1E-7)  & \hspace{0.55cm}26.84  & \hspace{0.4cm}26.02  & \hspace{0.4cm}26.73 \\

                          & Deep   &  & 46 & \hspace{0.4cm}$63232\ (64841)$        & 4.4E-6 (9.3E-8)        & 8.4E-6 (8.2E-8)      & \hspace{0.55cm}27.19            & \hspace{0.4cm}26.13  & \hspace{0.4cm}27.04 \\ \hline
\end{tabular}%
}
\end{table*}

\begin{figure*}
\includegraphics[width=0.9\textwidth]{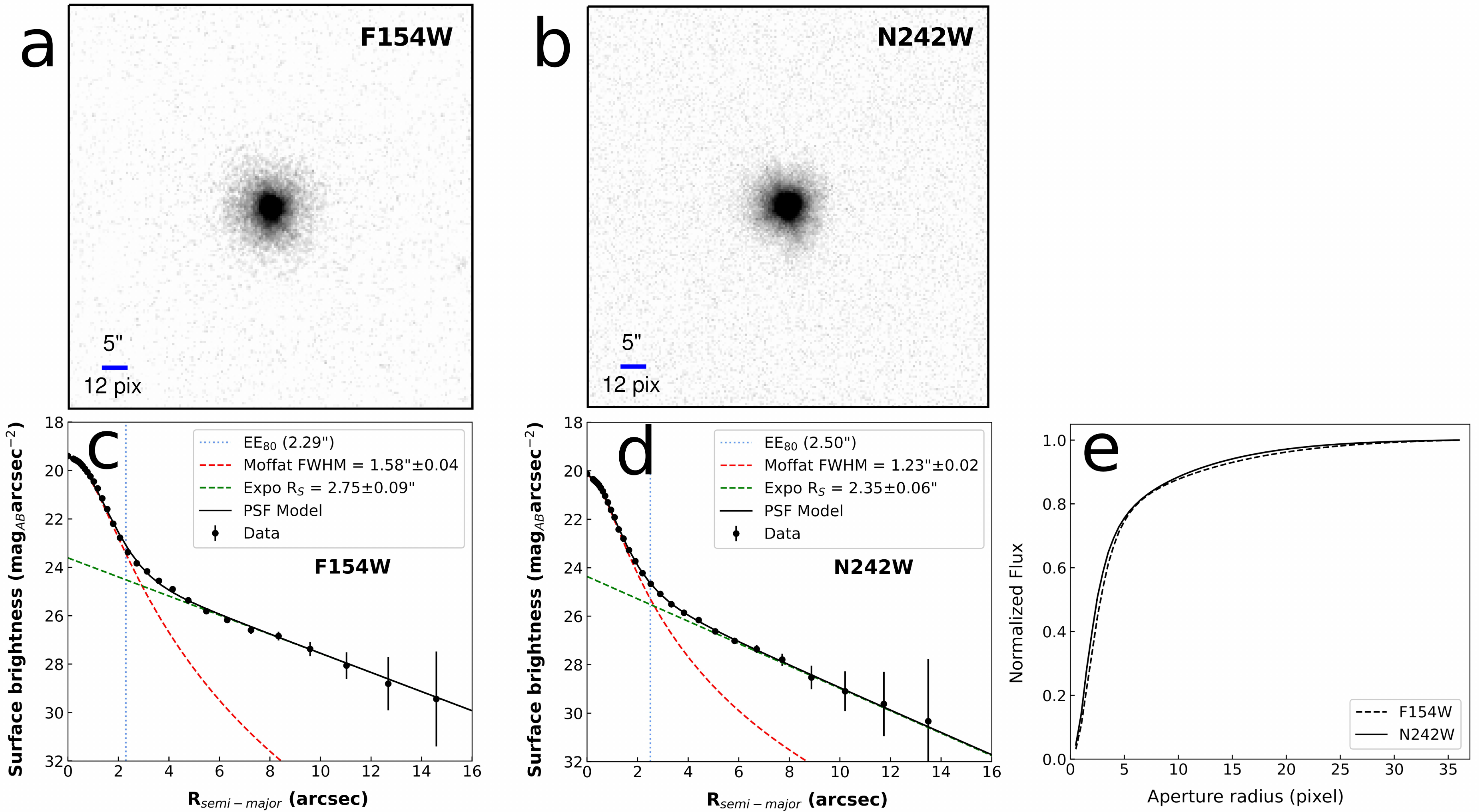}
\caption{Panel \textbf{a}: shows an image of a single star in the F154W filter. \textbf{b}: Mean stack of six stars in the N242W filter. \textbf{c,d}: Isophotal radial intensity profiles in F154W and N242W filters. The green and red lines show the individual moffat and exponential functions while the solid line represents the total fit to observed data points. The vertical dashed blue line indicates the radius of 80\% encircled energy (EE) on either panel. \textbf{e}: the normalized curve of growth for the F154W and N242W filters obtained using circular apertures.}
\label{fig:psf_fit}
\end{figure*}

\begin{table}
\caption{Percentage flux in F154W and N242W as obtained from COG in Figure \ref{fig:psf_fit}.}
\begin{tabular}{ccc}
\toprule
Aperture radius&  \%F154W &  \%N242W \\
 (pixel) & (0$\arcsec$.417 pix$^{-1}$) &  (0$\arcsec$.417 pix$^{-1}$)\\
\hline
1	&	10.56	&	14.06	\\
2	&	33.20	&	39.84	\\
3	&	53.47	&	58.44	\\
4	&	66.75	&	69.24	\\
5	&	74.58	&	75.58	\\
6	&	79.27	&	79.68	\\
7	&	82.32	&	82.63	\\
8	&	84.53	&	84.94	\\
9	&	86.27	&	86.84	\\
10	&	87.73	&	88.48	\\
11	&	89.02	&	89.90	\\
12	&	90.16	&	91.15	\\
13	&	91.20	&	92.26	\\
14	&	92.14	&	93.24	\\
15	&	93.00	&	94.11	\\
16	&	93.78	&	94.88	\\
17	&	94.49	&	95.57	\\
18	&	95.14	&	96.17	\\
19	&	95.72	&	96.70	\\
20	&	96.25	&	97.16	\\
21	&	96.72	&	97.57	\\
22	&	97.15	&	97.93	\\
23	&	97.54	&	98.25	\\
24	&	97.88	&	98.52	\\

\hline
\end{tabular}

\label{table:cog}
\end{table}

\begin{table}
\caption{Aperture correction values for some frequently used apertures obtained using COG.}
\centering
\begin{tabular}{ccc}
\toprule
Aperture radius&  $\Delta$m$_\textrm{FUV}$ &  $\Delta$m$_\textrm{NUV}$ \\
(in arcsec) &  &  \\
\hline
0.7 &   1.45    &   1.21 \\
1.0 &   0.94    &   0.79 \\
1.2 &   0.72    &   0.61 \\
1.4 &   0.57    &   0.50 \\
1.6 &   0.46    &   0.42 \\
\hline
\end{tabular}
    
\label{table:aper_corr}
\end{table}

\section{Point Spread Function and its pedestal}
\label{sec:PSF} 

\noindent We select a few stars in the field from the GAIA source catalog \citep{GaiaEDR32021} to characterize the UVIT PSFs. These stars are carefully chosen such that they are free from any contamination near the core and relatively brighter in the F154W and N242W band. While selecting these stars, we also avoided any double stars. Within the GOODS-South region (as shown in Figure~\ref{fig:rgb}), there is only one good star with F154W counterpart (with $17.8$ AB mag) and six stars ($19.5 - 18.4$~AB mag) in the N242W band, see Figure~\ref{fig:psf_fit} (a, b). Intervening objects are masked using SExtractor and the masked regions are filled with Gaussian noise matching the surrounding sky area. Any residual contamination is further removed by visual inspection. A mean stack is used to reduce noise and mitigate outliers while stacking the N242W stars. While preparing the stack, we have also checked a median stack of the stars and found that the 80\% encircled energy radius is larger by $\sim 0.2\arcsec$. Keeping this in mind, we chose the mean stack which should be a better representation of the empirical PSF. Besides, while working in the Poissonian data regime (where $\tilde{B} \times t_\textrm{exp} =\lambda \le 10$), opting for a median stack of the stars would result in an extra uncertainty of least count ($\sim 1/t_\textrm{exp} $ cps/pixel), which will be significant in the faint PSF wings where the flux in the wings is similar to the background noise $\sim \mathcal{O}(\sqrt{\lambda}/t_\textrm{exp})$.  However, this uncertainty is lower when choosing mean stacking, where it becomes $\sim 1/nt_\textrm{exp}$ (where n is number of stars choosen). Prior to stacking, we apply the saturation correction \citep{Tandonetal2020, Devrajetal2023} to all the stars having a total flux $\gtrsim$ 1 cps within a radius of $5\arcsec$. If the photon count is greater than 3 cps, there is more than 5\% saturation within a radius of $5\arcsec$ and that would affect the measured FWHM of the PSF. 
Bright point sources that are representative of the instrumental PSF can usually be dissected into a core and an extended wing or pedestal. To model the PSF in each filter, we use a combination of a Moffat \citep{moffat1969} for the core and an exponential profile \citep{Sahaetal2021,Borgohainetal2022} for the wing, given by the following relation:

 \begin{equation}
     PSF(r)= I_\textrm{c0} \left[ 1+\left(\frac{r}{\alpha}\right)^2 \right]^{-\beta} + I_\textrm{w0}\, e^{-r/{R_\textrm{s}}},
\label{eq:e_psf}
\end{equation}

 \noindent where the first term represents the Moffat component and symbols have their standard meaning; $R_\textrm{s}$ denotes the scale-length of the exponential profile, $I_\textrm{w0}$ is the extrapolated central intensity of the wing component. We use the IRAF \citep{tody1986} ELLIPSE task \citep{Jedrzejewski87} to obtain 1D intensity profile along the fitted isophotal ellipse semi-major axes. The observed intensity profile is then fitted with the model (Eq.~\ref{eq:e_psf}). We first obtain the scale-length of the wing by fitting an exponential to the PSF profile from $5 - 15
\arcsec$ as the saturation of the core does not significantly affect the region beyond $5\arcsec$. Note that $5\arcsec$ is also the radius beyond which the slope of the observed surface brightness profile follows a linear relation with the semi-major axis. The scale length obtained is then used as a fixed parameter while we fit the full profile with a combination of Moffat and exponential functions. As presented in Figure~\ref{fig:psf_fit} (c, d), we obtain the Moffat parameters to be $\alpha_\textrm{FUV} = 1\arcsec.71, \ \beta_\textrm{FUV} = 3.58$ , $\alpha_\textrm{NUV} = 1\arcsec.13, \ \beta_\textrm{NUV} = 2.66$. Then using the following relation: 
 
\begin{equation}
     FWHM = 2\alpha\sqrt{2^{1/\beta}-1},
\end{equation}
 
\noindent we obtain an FWHM of 1$\arcsec$.58 and 1$\arcsec$.23 for F154W and N242W respectively. The fitted exponential scale lengths to the extended wings of the PSFs come out to be 2$\arcsec$.75 and 2$\arcsec$.35 for F154W and N242W respectively. In Figure~\ref{fig:psf_fit}~(e), we show the curve of growth (COG) extracted using circular apertures on the modelled PSF, which is a good approximation as both UVIT PSFs are nearly circular. The COGs have been normalized to within $\sim$15\arcsec. The cumulative growth curve is found to practically flatten out at a radius $\sim$14-15\arcsec. Inspection of the COGs further out in the stamp also revealed that the total magnitude increased by a meagre $\sim$0.01 mag. Additionally, the surface brightness reaches the background sky level at around 14-15\arcsec radius, thereby making the normalization reasonable. No attempt has been made to characterize scattered light beyond the stamps due to higher RMS noise variations. These COGs are used to calculate the encircled energy and aperture correction. Table~\ref{table:aper_corr} lists a set of radii with aperture correction for both filters. From their COGs, we find that 80\% encircled energy (EE$_{80}$) corresponds to $\sim 2.3\arcsec$ and $\sim2.5\arcsec$ radii for the F154W and N242W filters respectively. The contribution of the wings in the measurement of total flux becomes important while measuring faint sources due to the difference between PSF FWHM and EE$_{80}$. The encircled energy outside 40 pixels (normalized to 90 pixels $\sim$37.5\arcsec) for the FUV PSF model is $\sim 0.37\%$ and for the mean stacked NUV PSF model, it is $\sim 0.16\%$. This number is less than what is obtained in the additional calibration paper by \cite{Tandonetal2020} for the UVIT PSFs ($\sim$2\% or 0.02 mag). It is important to note that in order to have a reliable measurement of this `small' flux, one needs a strong and isolated star as presented in \cite{Tandonetal2020}.

 
\section{Source detection, extraction, and photometry}
\label{sec:Sextractor}
We carry out source detection and photometry of objects in the field using the publicly available software, Source Extractor v2.5 \citep{Bertin96}. The detection of sources in the FUV and NUV wavelength regime is a complex process, especially due to the low photon statistics (Poisson nature). In addition, as these photons might have been emitted only from a certain part of a star-forming galaxy, it is often hard to identify or associate them with an optical counterpart; in other words, identification of galaxies based on morphological resemblance becomes a challenging task. The basic detection parameters used for the source detection are specified in Table~\ref{tab:sep_table}. The choice of these parameters is driven by the motivation to detect faint sources in the field with reliable photometric measurements limited to $\sim 3\sigma$ detection limit defined within a circular aperture by -

\begin{equation}
    {\widetilde{m}}_{3\sigma} = -2.5\log[{3\sqrt{\pi}{\widetilde{\sigma_\textrm{s}}}\times r}] + ZP_{\lambda}.
\end{equation}

\noindent Here, the circular aperture of radius $r$ (in pixel unit) is chosen to be the PSF FWHM for the given filter image and $ZP_{\lambda}$ is the magnitude zero-point in that filter. The parameter $\widetilde{\sigma_\textrm{s}}$ refers to the global standard deviation of the background (or simply rms). For both UVIT bands, the detection thresholds are relative to pixel values in the background rms maps that we provide while running SExtractor on a background-subtracted image. The use of the background rms map alleviates the detection of noise peaks or spurious sources in the low SNR areas of the image. For a relatively crowded field or a patch with high source density, the confusion noise limits the detection \citep{Morrisseyetal2007}. 

\begin{table}[]
\centering
 \caption{Source Extractor parameters used for source detection in the UVIT F154W and N242W filters. \label{tab:sep_table}}
\resizebox{0.5\textwidth}{!}{\begin{tabular}{ccc}
\hline
\multicolumn{1}{c}{Parameter name} & \multicolumn{1}{c}{\begin{tabular}[c]{@{}c@{}}\hspace{-1cm}Value \\ \hspace{-1cm}(F154W)\end{tabular}} & \multicolumn{1}{c}{\begin{tabular}[c]{@{}c@{}}\hspace{-1cm}Value\\ \hspace{-1cm}(N242W)\end{tabular}} \\ \hline
DETECT\_MINAREA                            & 11                                                                                             & 7                                                                                            \\
THRESH\_TYPE                            & RELATIVE                                                                                             & RELATIVE                                                                                            \\
DETECT\_THRESH                               & 0.8                                                                                           & 1 (hot), 3 (cold)                                                                                            \\

FILTER\_NAME                       & gauss\_1.5\_3x3.conv                                                                                  & gauss\_1.2\_3x3.conv                                                                           \\
DEBLEND\_NTHRESH                     & 32                                                                                            & 32                                                                                           \\
DEBLEND\_MINCONT                        & 0.0005                                                                                        & 0.001 
\\
CLEAN\_PARAM                        & 1                                                                                        & 1 
\\
WEIGHT\_TYPE                        & MAP\_RMS                                                                                        & MAP\_RMS
\\
PHOT\_APERTURES                        & 3.36,5.75,7.67                                                                                        & 3.36,5.75,7.67
\\
PHOT\_AUTOPARAMS                        & 2.5,3.5                                                                                        & 2.5,3.5
\\
PIXEL\_SCALE                        & 0.417                                                                                       & 0.417
\\
BACK\_TYPE                        & MANUAL                                                                                    & MANUAL
\\
BACK\_VALUE                        & 0                                                                                    & 0
\end{tabular}}

\end{table}

\begin{figure}
\includegraphics[width=0.5\textwidth]{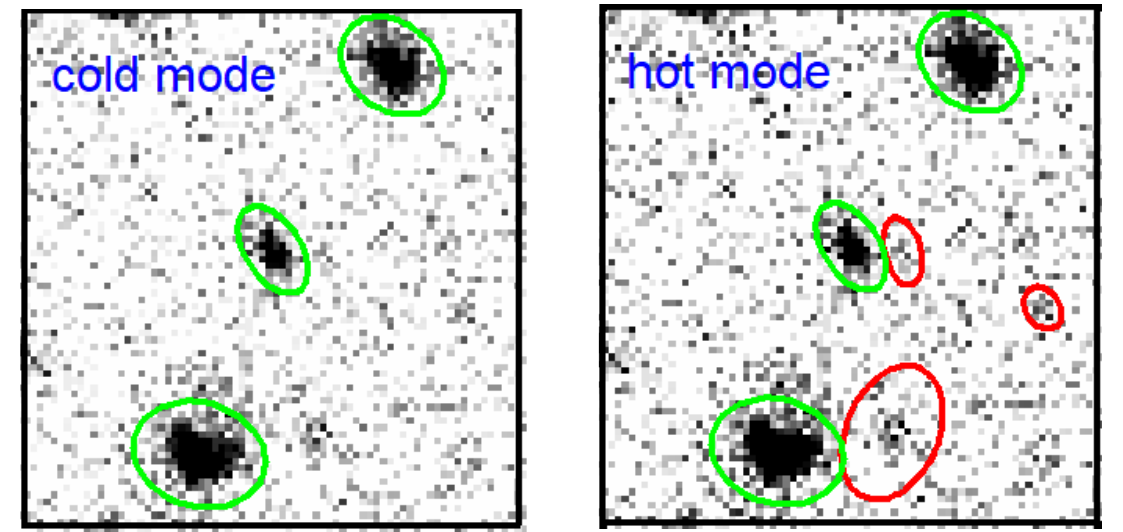}
\caption{Hot and cold mode detection in the N242W image. Kron ellipses (green - cold; red - hot) denote some of the detected sources.}
\label{fig:hotcold}
\end{figure}

\begin{figure*}
\begin{flushleft}
\includegraphics[width=1\textwidth]{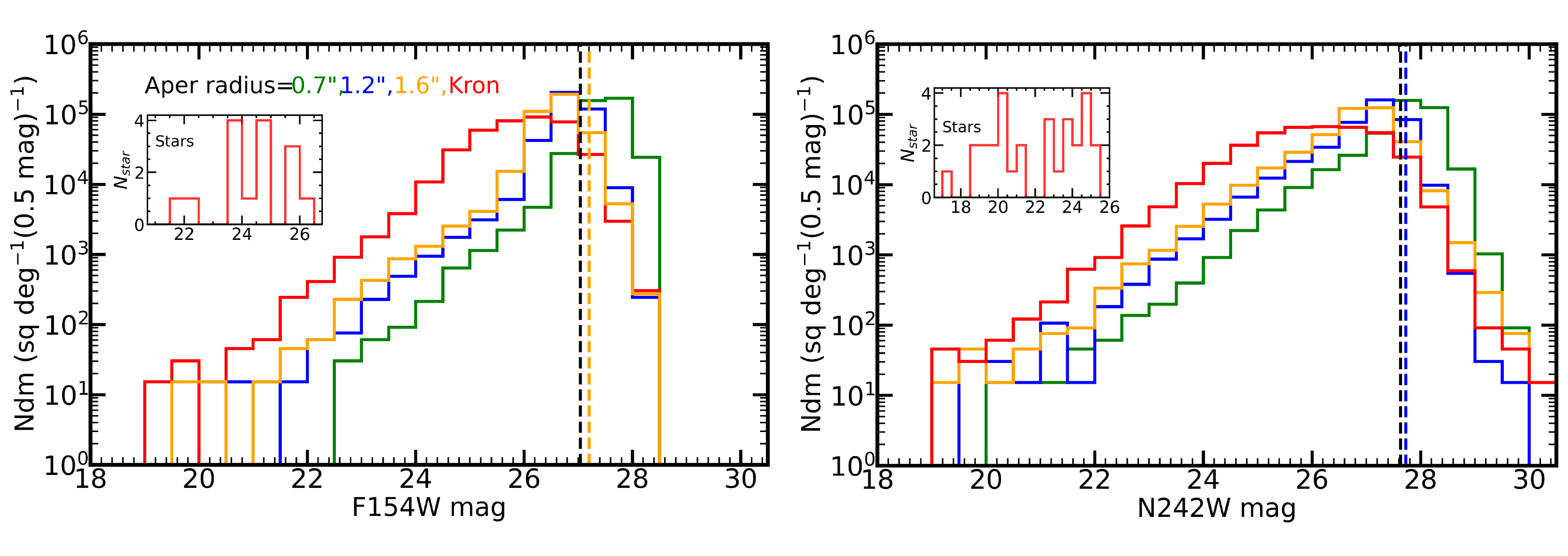} 
\caption{The differential number count (number/$\text{deg}^{2}/0.5mag$) of all objects (minus the stars) detected by SExtractor in the AUDFs field using the detection parameters in Table \ref{tab:sep_table}  (left: F154W, right: N242W). The inset plots denote the kron magnitude distribution of identified stars in the catalog. The source magnitudes are measured within apertures of radius 0.7\arcsec (green), 1.2\arcsec (blue), 1.6\arcsec (golden) and the kron radius (red). The $3\sigma$ detection limits are marked by golden dashed line (F154W) and blue dashed line (N242W). Also indicated by the black dashed lines are the 50$\%$ completeness limits.}
\label{fig:mag_distbn_FUV_NUV}
\end{flushleft}
\end{figure*}

While we employ the background rms map created in section~\ref{sec:background} throughout the detection run, the rest of the SExtractor parameters are determined by our experience gained in the catalog preparation for the AUDF north field \citep{Mondal23} as well as handling the current field (which brings out an added layer of complexity due to deep and shallow exposures). In Appendix~\ref{sec:sex_param}, we discuss how we choose an optimal set of input parameters for the SExtractor run such convolution filter, $DETECT\_MINAREA$, threshold. For the N242W image, where the FWHM of the Gaussian kernel is 1.2 pixels, we use the appropriate convolution filter $gauss\_1.2\_3x3.conv$. For the F154W image,we decided to use the appropriate convolution filter $gauss\_1.5\_3x3.conv$.

Besides the filter choice, other parameters such as the $Threshold$, $DETECT\_MINAREA$, deblending parameters are also known to strongly affect the source detection when large dynamic range is expected in the source fluxes (typical in deep field imaging) as well as their apparent sizes on the sky. For example, our experience suggests that a single set of $Threshold$ and deblending parameters might lead to a faint source blended with a nearby bright source, a big galaxy segmented into multiple components, or two physically disconnected sources to have the same segmentation number. 
To mitigate this issue and minimize using multiple sets of parameters, for the N242W image, we employ SExtractor in two different configurations - namely the hot and cold mode detection \citep{Rixetal2004,bardenetal2012,Rafelskietal2015}. In the cold mode (typically, with a high detection threshold, $\sim 3\sigma$, where $\sigma$ is estimated from the background rms map), spurious deblending of large sources with clumps, spiral structures or any substructures is avoided but fails to detect faint, low-surface brightness (LSB) galaxies (see Figure ~\ref{fig:hotcold} for an illustration). On the other hand, the hot mode (typically, with a lower detection threshold, $\sim 1\sigma$) setting is ideal for detecting faint sources and minimizing the detection of spurious sources or noise peaks. In either case (i.e., hot and cold mode), a source is defined to have a minimum of 7 contiguous pixels (equivalent to a circular aperture with diameter=FWHM of the N242W PSF) above the detection threshold. The rest of the parameters of the cold and hot mode are presented in Table~\ref{tab:sep_table}. Note that the two-step detection process does not affect the detection of sources brighter than 27 AB mag in the N242W image. After carrying out the hot and cold mode detection, the identified sources are cross-matched with each other, to prepare a combined catalog with a unique identification number for detected objects. Similar to the approach of \cite{bardenetal2012}, hot mode objects whose centers lie within the \textit{Kron ellipse} of a cold mode detected object are removed from the final list of detected objects. As an example, in the left panel of Figure \ref{fig:hotcold}, if we consider the lowermost cold mode detected object (within the Green Kron aperture), this object is also detected in the hot mode run (right panel). However, clumps surrounding the object, which are visually identifiable and are segmented in the hot mode are excluded in the final list of detected objects. Note that even with this two-step detection process, some of the genuine faint sources might remain missing and this will be investigated in a future paper.  

For the F154W image, however, we restricted ourselves to only a single detection mode.  To avoid detecting spurious, faint sources, we increased $DETECT\_MINAREA =11$ (equal to a circular aperture of diameter=FWHM of F154W PSF), while we reduced the $Threshold=0.8\sigma$ to redefine an FUV source.  
All other parameters are presented in Table~\ref{tab:sep_table}. For both images, the clean parameter has been kept fixed at 1 for all catalogs to avoid unwanted (or unreal) faint sources around bright objects.  

\defcitealias{Whittakeretal2019}{Wh19}

\begin{figure}
\begin{center}
\rotatebox{0}{\includegraphics[width=0.5\textwidth]{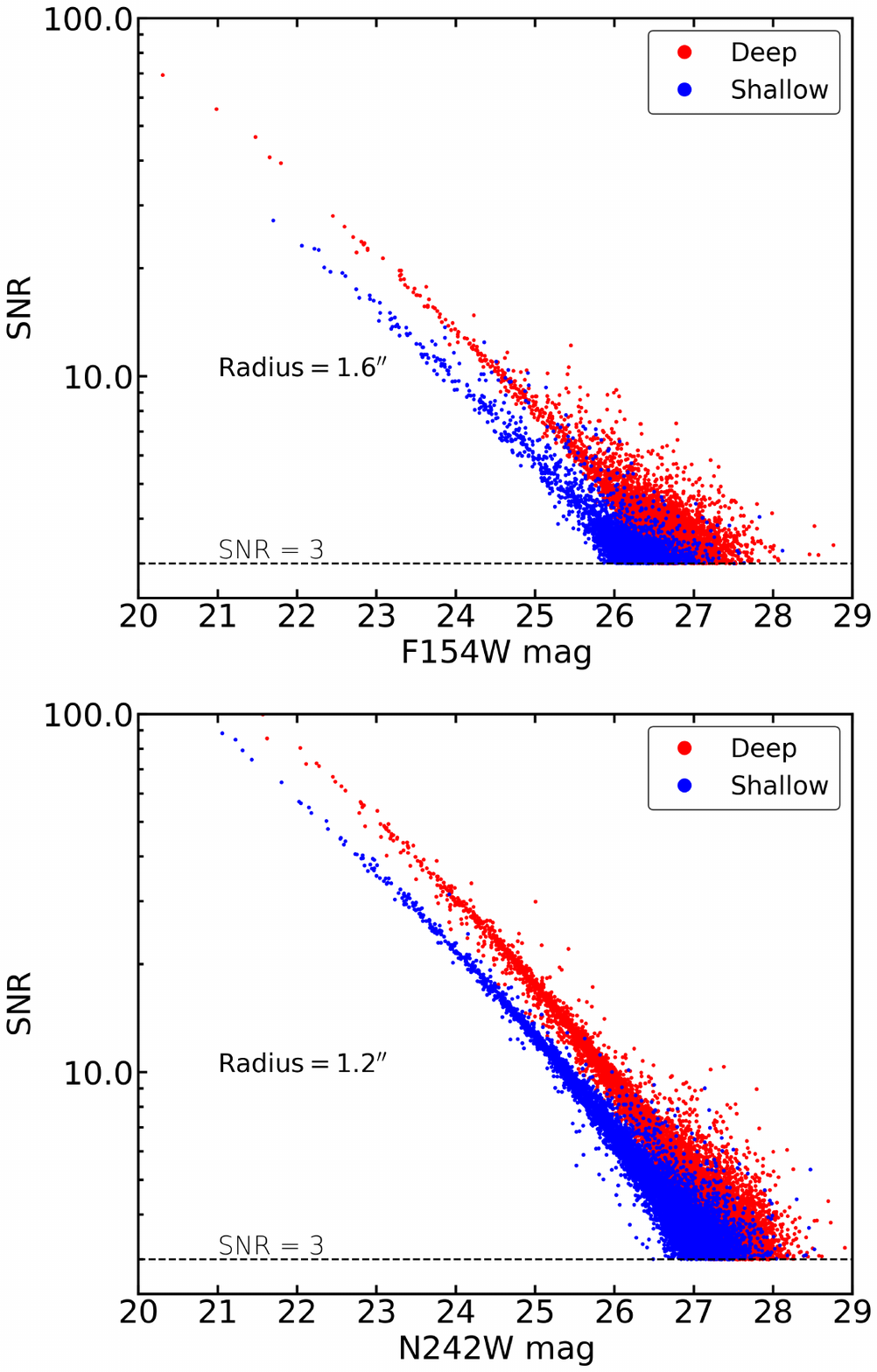}}
\caption{Magnitude v/s SNR plot for the F154W (Upper Panel) and N242W (Lower Panel) objects in the UVIT catalogs showing the effect of exposure time in deep and shallow regions of the image. The effect is prominent for relatively brighter sources.}
\label{fig:deep_shallow_snr}
\end{center}
\end{figure}

We include further details relating to the susceptibility to identifying noise peaks as credible objects using the aforementioned detection parameters in Appendix \ref{sec:poisson_exp}. Our definition of a noise peak is an object that is formed out of pure statistical fluctuations in the background, especially due to the chance alignment of background noise pixels and detected by SExtractor. In the Appendix \ref{sec:poisson_exp}, we carry out a detailed experiment to detect and quantify the fraction of such sources (i.e., noise peaks) in our final image. From Table~\ref{tab:poisson_exp_table}, it can be seen that at a magnitude brighter than 27 AB mag, there are no credible noise peaks in the field. In the magnitude range $27 - 27.5$ mag, there are $\sim 0.2\%$ and $\sim 7\%$ noise peaks in N242W and F154W images respectively. The noise peak detection starts rising for magnitudes fainter than the $3\sigma$ detection limit.

\begin{figure*}
\centering
\includegraphics[width=1.7\columnwidth]{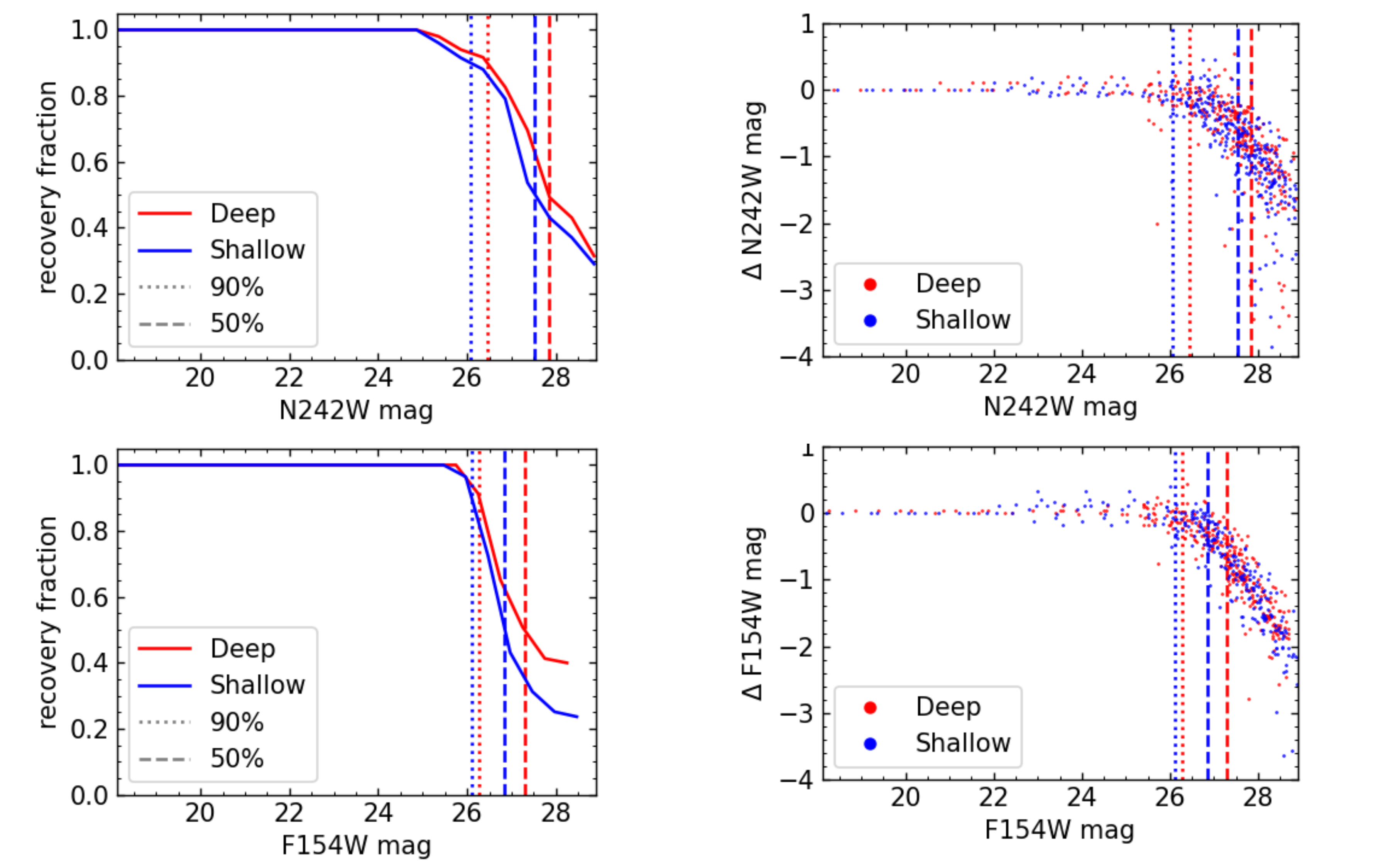}
\caption{[Top Left] The recovery fraction of injected simulated sources as a function of input magnitude (aperture radius=1.2$\arcsec$) for the N242W image region. The deep and shallow image recovery fractions are marked in red and blue respectively. The 90\% and 50\% completeness limits for the Deep N42W image region are marked as dotted and dashed lines respectively. [Top right] The offset of measured magnitude from the input magnitude (aperture radius=1.2$\arcsec$) of injected simulated sources as a function of the same N242W input magnitude. [Bottom] Same as top panel but for the F154W image regions (aperture radius=1.6$\arcsec$).}
\label{fig:artificial}
\end{figure*}

\subsection{Photometric catalogs and source density}
\label{sec:catalog}
We carry out the photometric measurements of detected objects using SExtractor within fixed circular apertures, and Kron radii \citep{Kron1980}. We show the differential source count distribution of all sources detected in the field by SExtractor in Figure ~\ref{fig:mag_distbn_FUV_NUV}. The inset plots denote the distribution of stars in the field, that have been cross-matched with GAIA DR3 \citep{GaiaEDR32021} catalog and this distribution is more or less flat. All magnitudes can be corrected using the curve of growth (Figure~\ref{fig:psf_fit}) and Table~\ref{table:cog}. However, for large extended objects, the growth curve corrections would not be appropriate; hence, we also include the Kron magnitudes of objects in the UVIT catalogs. We estimate the signal-to-noise for all the detected objects using the following equation-

\begin{equation}
    SNR = \frac{S\times \sqrt{t_\textrm{exp}}} {\sqrt{S + {\widetilde{B_\textrm{s}}\times \pi r^2}}}, 
    \label{eq:SN}
\end{equation}

\noindent where S denotes the background-subtracted source flux within a circular aperture of radius $r$, $\widetilde{B_\textrm{s}}$ is calculated using the background map, and $t_\textrm{exp}$ is the exposure time for the source. While writing the noise term in the denominator of the above equation, we have explicitly neglected read noise that is negligible in the CMOS detector (where each pixel contains a photodiode and a current amplifier) used in UVIT. The dark current is $\sim 7.8\times 10^{-7}$~cps~pix$^{-1}$ which is equivalent to $10 e^{-1}$~s$^{-1}$ for the full FOV (i.e., 14 arcmin radius). The background estimated (Figure~\ref{fig:backmap}) above includes a contribution from the dark current and is much smaller than the actual sky background \citep[see also][]{Sahaetal2020}. For estimating the SNR, we carried out forced photometry using FWHM-sized apertures of radii $1.6"$ for F154W and $1.2"$ for N242W. The SNR vs the magnitude of detected objects is shown in Figure \ref{fig:deep_shallow_snr}. Two distinct branches, marked by red and blue dots are visible. These arise due to the difference in the exposure time in the regions (deep and shallow) where these objects are detected. The current UVIT catalogs contain only those sources that have an $SNR \ge 3$. We have listed these signal-to-noise estimates under the \textit{snr column} in the UVIT catalogs (Table \ref{tab:UVIT_cat}). 
For every object in the F154W (N242W) catalog, we have indicated the presence of an object in the N242W (F154W) catalog within a distance of PSF FWHM $\sim 1.6\arcsec (1.2\arcsec)$ from the object using a flag. For example, we assign F154W\_flag =1 when there is an object in the N242W catalog and F154W\_flag =0 when there is none. Similarly, for the N242W\_flag. In this work, we present the following catalogs: \\

\noindent {\it AUDFs-F154W-Cat0-v1 }: F154W catalog consisting of detected UVIT objects with SNR greater than 3; {v1} refers to version 1. 

\noindent {\it AUDFs-N242W-Cat0-v1}: N242W catalog consisting of detected UVIT objects with SNR greater than 3; {v1} refers to version 1. 

\noindent {\it AUDFs-N242W-Cat1-v1}: N242W cross-matched catalog comprising of those objects in {\it AUDFs-N242W-Cat0-v1} having an individual or multiple counterparts in the photometric catalog of \cite{Whittakeretal2019}. Also provided are some information regarding the candidate optical(HST)/FUV(F154W) counterparts for each object in this catalog (See Section \ref{sec:hst_crossmatch}); {v1} refers to version 1.

\noindent The content of the UVIT catalogs is provided in Table \ref{tab:UVIT_cat}. In addition to the detector, sky coordinates and all the aperture (e.g., 0.7\arcsec, 1.2\arcsec, 1.6\arcsec) magnitudes, kron magnitudes and their error, we also present some basic structural parameters of the sources such as their semi-major and semi-minor axes, position angle on the sky. The catalog magnitudes are not foreground dust corrected. However, the Galactic extinction can be corrected following the Galactic extinction law with $R_\textrm{V} =3.1$ \citep{Fitzpatrick1999} with the average extinction towards GOODS South being $A_\textrm{V}=0.0248$ mag \citep{Schlegeletal1998}. This translates to $A_\textrm{F154W}=0.063$ and $A_\textrm{N242W}=0.061$ mag for F154W and N242W respectively. 

\begin{table}
\centering
\caption{Content of \textit{AUDFs-F154W-Cat0-v1} and \textit{AUDFs-N242W-Cat0-v1}. Note the radii (in pixels) of the semi-major and semi-minor axis of the Kron aperture are obtained by multiplying the kron radius factor with the a\_image and b\_image parameters respectively.}
\label{tab:UVIT_cat}
\begin{tabular}{p{2.5cm}p{5.5cm}}
\hline
Column name & Description\\\hline

UVIT\_ID & Identification number in the catalog\\
SOURCE\_NAME & Unique string associated with object\\

X                                                                     & X pixel coordinate of the object centroid\\
Y                                                                     & Y pixel coordinate of the object centroid\\
RA & RA (J2000) of the object's centroid\\
DEC & Dec (J2000) of the object's centroid\\
UVIT\_m                                                               & Magnitude within an aperture of \\
          &radius 0.7"(m1), 1.2"(m2) and 1.6"(m3)  \\
UVIT\_merr                                                            & Error in magnitude                                                                                                  \\
UVIT\_AUTO                                                            & Magnitude within the kron aperture                                                                                            \\
UVIT\_AUTO\_err                                                       & Error in magnitude within the kron aperture                                                                                                      \\
flux\_radius                                                          & radius enclosing 50 percent of \\ & the object's light (in pixels)                  \\
kron\_radius$^{*}$                                                          & {Reduced Kron Pseudoradius factor}                                                                                                                 \\

a\_image                                                              & Semi-major axis of the object (in pixels)                                                                                                        \\
a\_image\_err                                                         & Error in a\_image                                                                                                                                \\
b\_image                                                              & Semi-minor axis of the object (in pixels)                                                                                                        \\
b\_image\_err                                                         & Error in b\_image                                                                                                                                \\
PA                                                                    & Position angle (in degrees)                                                                                                                     \\
SNR & Signal-to-noise estimated using Eq \ref{eq:SN} \\

F154W\_flag & presence of an object in the N242W catalog   \\

N242W\_flag &  presence of an object in the F154W catalog  \\
\end{tabular}
\end{table}

\begin{table}
\centering
\caption{Content of \textit{AUDFs-N242W-Cat1-v1}.}
\label{tab:N242W_crosscat}
\begin{tabular}{p{3.0cm}p{5.5cm}}
\hline
Column name & Description\\\hline

N242W\_ID                                                              & Identification number (in \textit{AUDFs-N242W-Cat0-v1})                                                                                                                                  \\
N242W\_NAME                                                              & Source name (in \textit{AUDFs-N242W-Cat0-v1} )                                                                                                                                  \\
RA                                                                    & RA (J2000) of the object centroid                                                                                                               \\
Dec                                                                   & Dec (J2000) of the object centroid                                                                                                               \\
N242W\_m2                                                               & From \textit{AUDFs-N242W-Cat0-v1} \\
N242W\_m2err                                                            & From \textit{AUDFs-N242W-Cat0-v1}                                                                                                 \\
N242W\_AUTO                                                            & From \textit{AUDFs-N242W-Cat0-v1}                                                                                            \\
N242W\_AUTO\_err                                                       & From \textit{AUDFs-N242W-Cat0-v1}                                                   
\\
CANDELS\_ID\_nn                                                          & CANDELS\_ID of the nearest neighbour \\
             & from the \citetalias{Whittakeretal2019} catalog (see text) \\
CANDELS\_RA\_nn                                                           & RA (J2000) of CANDELS\_ID\_nn                                                                                 \\
CANDELS\_Dec\_nn                                                          & Dec (J2000) of CANDELS\_ID\_nn\\

F154W\_ID\_nn & F154W\_ID of the nearest neighbour \\
             & in the \textit{AUDFs-F154W-Cat0-v1} \\

F154W\_m3\_nn                                                               & F154W\_m3 from \textit{AUDFs-F154W-Cat0-v1}\\
F154W\_m3err\_nn                                                            & F154W\_m3err from \textit{AUDFs-F154W-Cat0-v1}                                                                                                \\
F154W\_AUTO\_nn                                                            & F154W\_AUTO from \textit{AUDFs-F154W-Cat0-v1}                                                                      \\
F154W\_AUTO\_err\_nn                                                       & F154W\_AUTO\_err from \textit{AUDFs-F154W-Cat0-v1}                                                                                                                             \\
CANDELS\_flag                                                         & Flag for the CANDELS counterpart                            \\

F154W\_flag  & Flag for the F154W counterpart   \\

\end{tabular}
\end{table}

Based on the current SExtractor parameters and methodology, we have a total of $20097$ and $13885$ objects in the N242W and F154W catalogs, respectively. Of these, there are $19374$ sources brighter than $3\sigma$ detection limit (27.7 AB mag) in the N242W band within an area of 236 sq arcmin on the sky.  While in the case of F154W filter, there are $13495$ objects brighter than $3\sigma$ detection limit (27.2 AB mag). Clearly, compared to the F154W filter, the source density is higher in the N242W filter. This is expected as not all galaxies would be forming stars in the past $100 - 200$~Myr (a timescale probed by stellar emission in the F154W band). Considering the circular aperture photometry (with aperture radius=1.2\arcsec), the galaxy count power-law slope measured in the flux ranges AB$\simeq 18 - 25 ~(26)$~mag is found to be 0.43 (0.46) dex~mag$^{-1}$ in our N242W filter. Our estimated N242W galaxy count slope closely matches 0.44 (0.43) dex~mag$^{-1}$ obtained using the HST/WFC3/UVIS F225W (F275W) filters in the magnitude range $19 - 25$~AB mag based on the ERS GOODS South observations \citep{Windhorstetal2011} as well as the UVIT/N242W slope of 0.44 dex~mag$^{-1}$ estimated on the GOODS North data in the flux range AB$\simeq 18 - 25$~mag \citep{Mondal23}. On the other hand, the GALEX NUV count slope 0.58 dex~mag$^{-1}$ of \cite{Xuetal2005} covering the fluxes in the range AB$\simeq 17 - 23$~mag ($\sim 2$~mag brighter) is higher compared to the WFC3/UVIS filters and our N242W filter slopes. In the case of the F154W filter, our measured slope 0.56 (0.57) dex~mag$^{-1}$ in the flux range AB$\simeq 20 - 25 ~(26)$~mag is consistent with previous estimates from GALEX FUV slope 0.51 dex~mag$^{-1}$ in the magnitude range $17 - 23$ AB mag in GOODS South \citep{Xuetal2005} and UVIT/F154W slope 0.57 dex~mag$^{-1}$ in the magnitude range $19 - 25$ AB mag in GOODS North \citep{Mondal23}. 

\subsection{Artificial source injection and Completeness} 
\label{sec:Completeness}
We carried out a completeness analysis of the UVIT AUDFs images through the injection and subsequent recovery of simulated point sources, as was done for the UVIT AUDFn images in \citet{Mondal23}, similar to the analysis carried out for optical and UV images in \citet{Tyson88}, \citet{Janes93}, \citet{Ferguson95}, \citet{teplitz2013} and more recently in \citet{Bhattacharya19}. The deep and shallow regions were treated separately for both  N242W and F154W images. Simulated sources of varying magnitude (each following the moffat PSF parameters described in Section~\ref{sec:PSF}; magnitude is then curve-of-growth corrected) were injected into the N242W and F154W images at randomised positions using the IRAF \textit{mkobjects} routine. Source detection and extraction were carried out as described earlier in Section~\ref{sec:Sextractor}. For each image, the recovery fraction is high for brighter sources and reduced for fainter sources. Figure~\ref{fig:artificial} shows the recovery fraction of injected simulated sources as a function of input magnitude for the deep and shallow regions of the N242W image (Top Left) and the F154W image regions (Bottom Left). 

The 90\% and 50\% completeness limits for the deep and shallow regions of each filter image are noted in Table~\ref{tab:back}. The increased exposure time of $\sim$2 times for the deep regions compared to the shallow regions is expected to allow detection of sources that are fainter by $\sim$0.4~mag in the deep images. This is reflected in the 0.35~mag (F154W) and 0.39~mag (N242W) differences between the 3$\sigma$ detection limits between the shallow and deep regions, and also in the 0.31~mag (F154W) and 0.33~mag (N242W) differences between their 50\% completeness limits. 

The completeness of our observation depends on several factors (including the exposure time, detector throughput, field source density and PSF) but most importantly on the source detection parameters described in Section~\ref{sec:Sextractor}. The detection parameters were so chosen to maximise detection of sources brighter than the 3$\sigma$-detection limit while also keeping spurious sources or noise peaks detection at bay (see Appendix~\ref{sec:poisson_exp}). The 50\% completeness limits being close to the 3$\sigma$-detection limit (Table~\ref{tab:back}) for both F154W \& N242W images, with negligible spurious sources or noise peaks (see Figure~\ref{fig:spurious_mag_distbn}), reflects the efficiency of our detection technique. There are 2341 and 13079 sources brighter than 90\% and 50\% completeness respectively in the F154W band. While in the N242W image, we have 7790 and 18824 objects brighter than 90\% and 50\% completeness.
We further calculate the offset of magnitudes measured by SExtractor (individual detection on each image) using the input magnitude from the simulated sources in each image. This offset is plotted in Figure~\ref{fig:artificial} as a function of input magnitude for the deep and shallow regions of the N242W (Top Right) and F154W (Bottom Right) images. The figures show that the recovered magnitude is close to the input value for all images for sources brighter than the 90\% completeness limit. The recovered magnitude is still close to the input value for sources brighter than the 50\% completeness limit, though the fainter sources in this range appear brighter than their input magnitudes. 

There are a number of sources, fainter than the 50\% completeness limit in F154W and N242W, that appear brighter than their input magnitude with the brightening increasing for fainter sources. Such flux boosting/Malmquist bias has long been observed in images with low photon counts when utilising aperture magnitudes and quantified using artificial star tests \citep[e.g.][]{Tyson88, Wang04}. As discussed in \citet{Mondal23}, since the source detection falls sharply beyond the 50\% detection limit there is a preference for only those sources brighter than their input values to be detected. Those few sources that are still detected likely have some background pixels above the mean, close to the source center. These are then counted as source pixels thereby increasing the source flux. Thus fainter sources appear brighter than their input magnitudes with this offset increasing with faintness, as seen in the right panels of Figure~\ref{fig:artificial}. The catalogs are presented in this work without correcting for flux boosting.

\begin{figure}
\hspace{-1.2cm}
\includegraphics[width=0.56\textwidth]{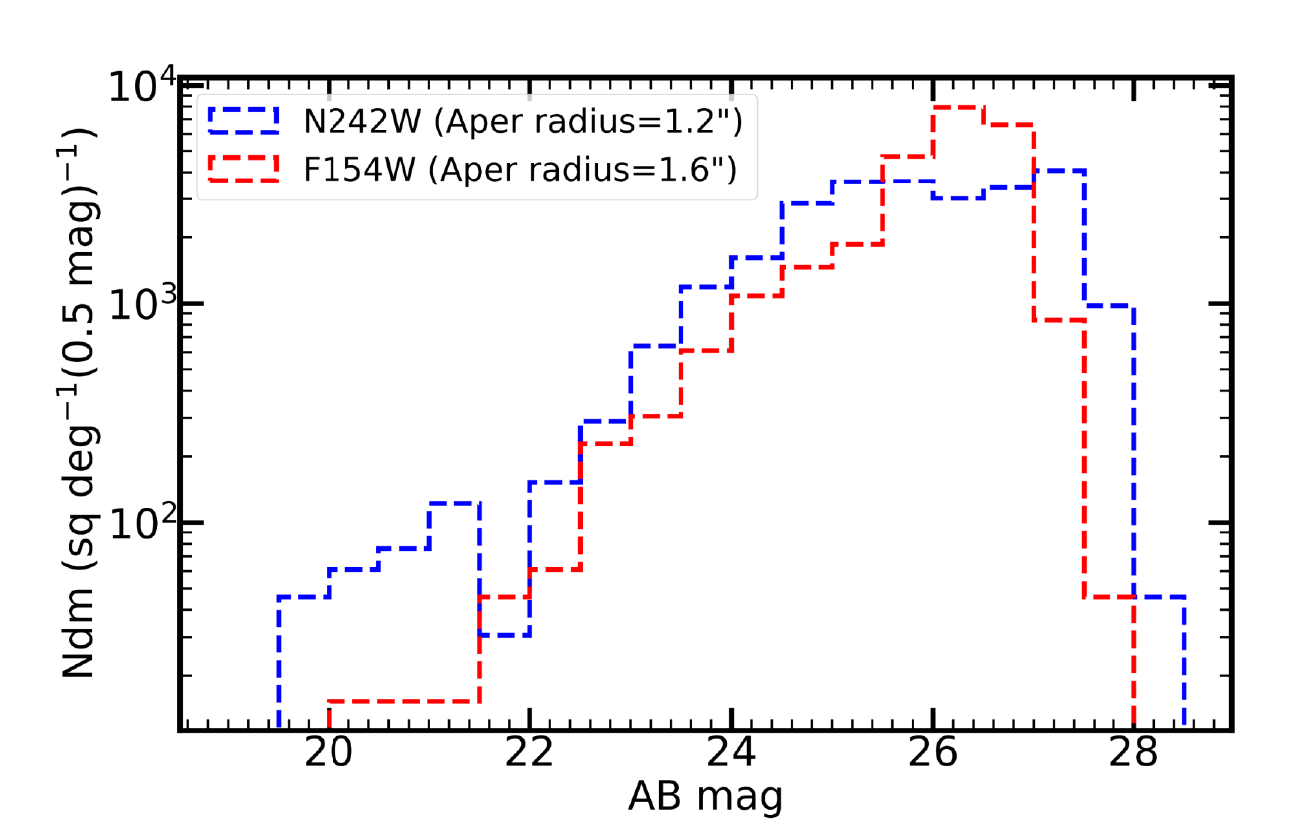}
\caption{Source count distribution (red: F154W, blue: N242W) of objects having a joint CANDELS and F154W\_flag$=1$ in the N242W-cross matched catalog. The measured source magnitudes are not aperture-corrected.}
\label{fig:unique_mag_distbn_FUV_NUV}
\end{figure}

\subsection{CANDELS optical/IR counterpart identification} \label{sec:hst_crossmatch}

The overlap of the selected field with the CANDELS GOODS-South (Figure~\ref{fig:rgb}) makes the identification of the objects in the UVIT catalogs with their optical/infrared counterparts in the CANDELS survey possible \citep{Guoetal2013}. This could be useful in several science cases, e.g., construction of the Spectral Energy Distribution, identifying possible Lyman-continuum leaker candidates with a given spectroscopic redshift, construction of the UV luminosity function or simply estimating the UV-optical and UV-IR color of an object. To aid the identification of possible optical/infrared counterparts in the CANDELS catalog, we provide \textit{AUDFs-N242W-Cat1-v1} catalog, with sources in the N242W catalog for which we have identified a likely HST optical/infrared counterpart in the HST-GOODS South photometric catalog (\cite{Whittakeretal2019}, hereafter \citetalias{Whittakeretal2019}). The flags in this catalog are determined by keeping in mind the differences in resolution between the UVIT and HST. We assign a CANDELS\_FLAG=1 to objects for which we are confident of identifying a unique counterpart in the HST catalog. These correspond to cases in which there is a single object in the \citetalias{Whittakeretal2019} catalog within a search radius of 1.2" from the center of the UVIT object in the N242W catalog. Note that the positional coordinates we use from the \citetalias{Whittakeretal2019} catalog for cross-matching correspond to the coordinates calibrated using GAIA astrometry denoted by '\textit{ra\_gaia}' and '\textit{dec\_gaia}'. The search radii correspond to the FWHM of the N242W PSF. For some bright extended sources, the distance between the optical and UVIT centers may be larger than 1.2/1.6". In such cases, we use the kron radius as the search radius instead of the PSF FWHM. We assign a CANDELS\_FLAG=2 to objects for which there is more than one object center in the HST catalog within the N242W PSF FWHM.
Similar to the identification of possible HST-detected counterparts to the N242W objects in this catalog, we have also provided similar information about possible far-ultraviolet counterparts in the F154W band by means of an F154W flag. Note, this F154W flag is slightly different than the one provided in Table \ref{tab:UVIT_cat}. While a value zero for the F154W flag still conveys that there is no object in the F154W catalog within the search radius of 1.2", a value 1, in this case, indicates that there is just one F154W detected object within the search radius. A value 2, similar to the CANDELS\_flag=2, indicates multiple F154W detected objects within the search radius. There are a total of 1695 objects in the N242W cross-matched catalog with a CANDELS and F154W\_flag=1. The source count distribution of these objects is shown in Figure \ref{fig:unique_mag_distbn_FUV_NUV}. A joint distribution of the difference in positions of bright sources in N242W and their unique counterparts in CANDELS is shown in Figure \ref{fig:diff_coord}. The source catalog of \citetalias{Whittakeretal2019} is constructed using the HST IR bands and hence the difference in the centroids essentially corresponds to differences in the centers of objects in different wavelength regimes. However, one needs to be careful here, since UV photons could arise from any part of a galaxy. We picked a few random sources (especially, the extended ones), estimated their centroid (center of light in X, Y by placing a box on the source) and found it to differ ($\sim 1\arcsec$ or so) from what is provided by SExtractor. However, this problem is not so severe for smaller-sized sources and large extended sources are not many in the F154W and N242W filters. The content of this catalog is provided in Table \ref{tab:N242W_crosscat}.

\begin{figure}
\centering
\includegraphics[width=0.5\textwidth]{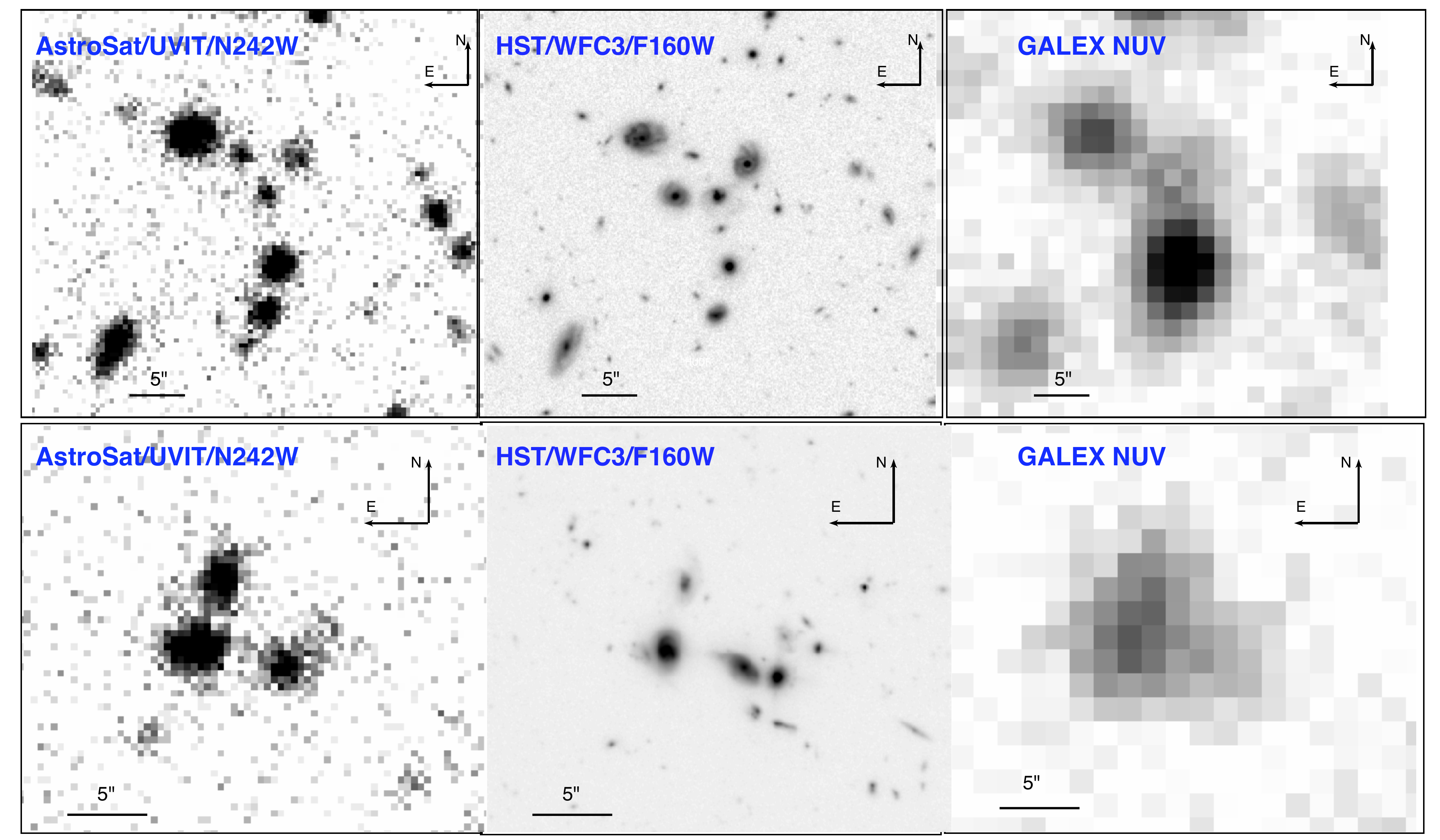}
\includegraphics[width=.5\textwidth]{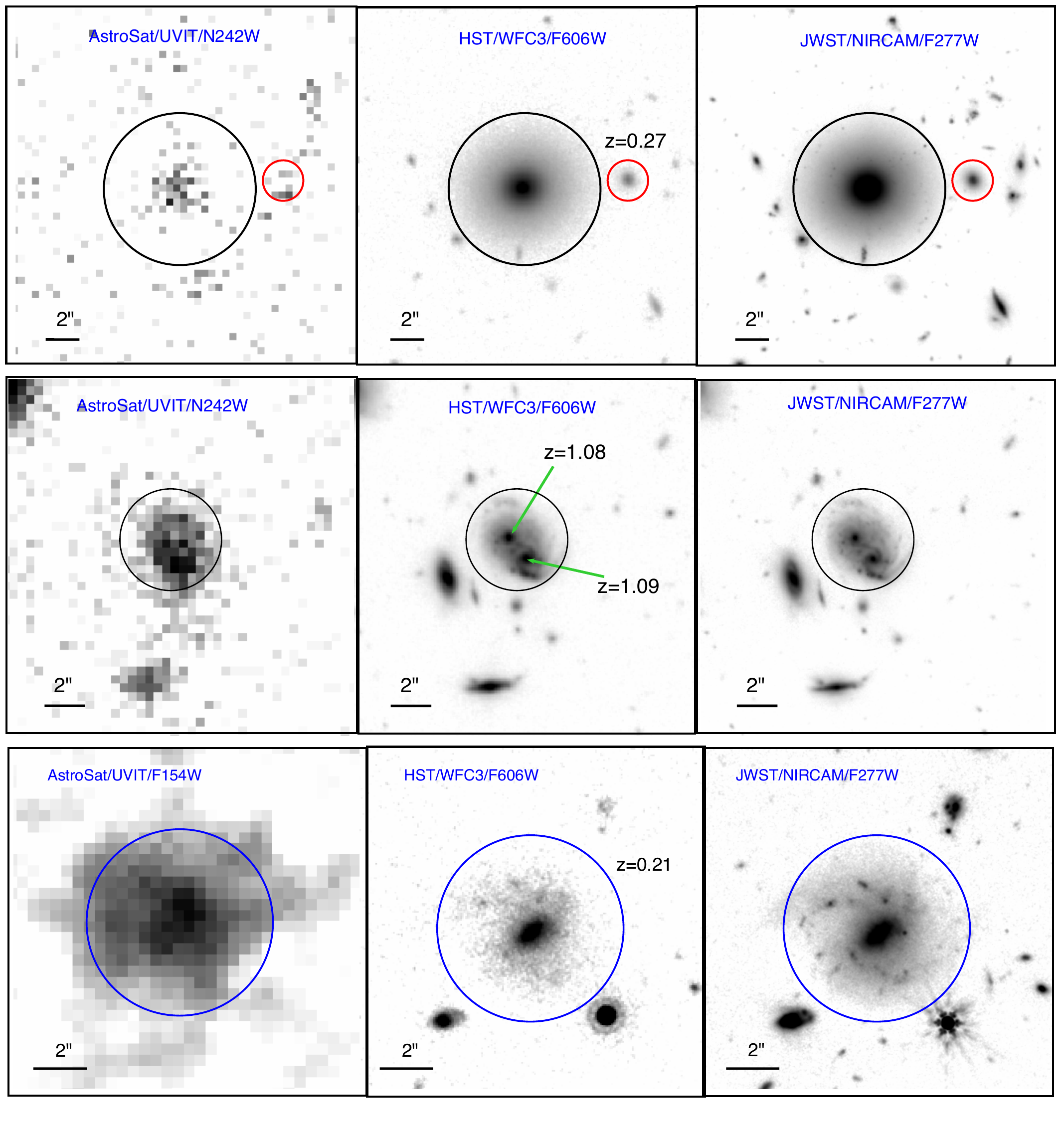}
\caption{Top two rows: A comparison of GALEX NUV and UVIT NUV observation with HST/F606W filter image as the optical reference. The third, fourth and fifth rows: central UV emission from a local Elliptical (E0) galaxy in N242W filter. HST/F606W and JWST/F277W images are shown to confirm the morphology; the radius of the black circle is 4.5\arcsec. The red circle denotes another early-type galaxy at z=0.27.
Middle row: an interacting spiral galaxies at $z \simeq 1$ in N242W filter, circle size 2.5\arcsec. HST/F606W and JWST/F277W images are shown for the sake of comparison. 
Bottom row: UVIT FUV image of a barred spiral galaxies at $z \simeq 0.21$; blue circle size 3.5\arcsec}
\label{fig:interactingspiral}
\end{figure}

\begin{figure}
\includegraphics[width=0.5\textwidth]{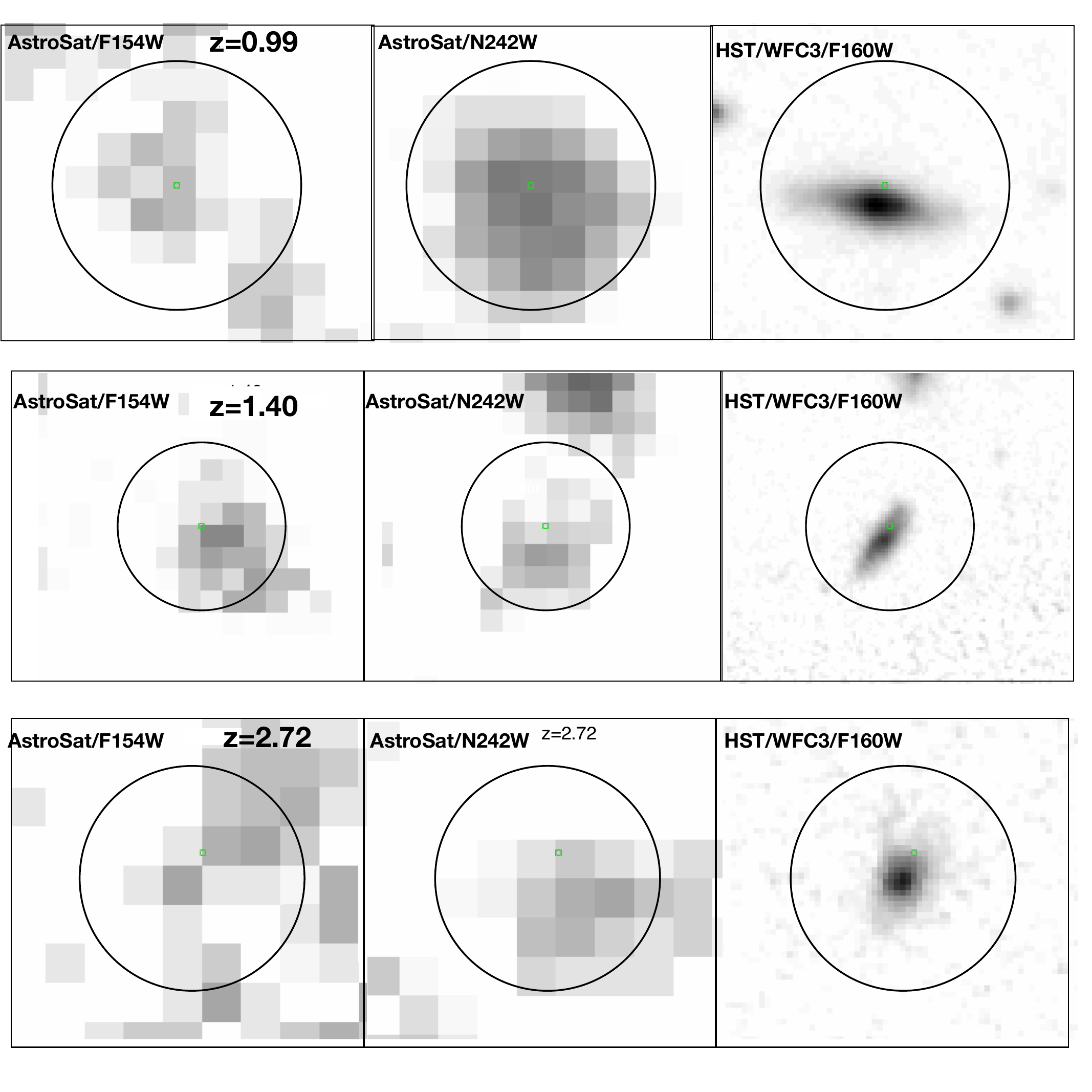}
\caption{Examples of candidate LyC leakers caught in F154W (upper and middle panels) and N242W (bottom panel) filters from AUDF South field. The radius of the dark circle in each panel is 1.6". The green square indicates the on-sky position of these galaxies taken from HST/G141 grism survey \citep{Momchevaetal2016}.}
\label{fig:LyC}
\end{figure}

\section{Interesting galaxy images from UVIT}
\label{sec:imagecomp}

Figure~\ref{fig:interactingspiral} displays a set of images for the sake of visual comparison with the same from other telescopes such as GALEX, HST, JWST. The first two rows showcase the significant improvement in the image quality and PSF (1.2\arcsec - 1.5\arcsec) of UVIT over GALEX (4\arcsec - 6\arcsec) - on either panel, one can see how three or more sources are nearly merged into one in GALEX while they can be clearly identified in UVIT as well as in the higher resolution HST/F160W band IR images. The better resolution of UVIT and astrometry have also reflected in the galaxy count slope being consistent with HST/WFC3/UVIS filter images (see Section~\ref{sec:catalog}). 

In the next three rows of Figure~\ref{fig:interactingspiral}, we illustrate a few interesting science cases that can be pursued using these imaging data. The first one is a typical bright elliptical (E0) galaxy seen in HST optical and near-IR images from JWST JADES survey \citep{Eisensteinetal2023b}. The N242W data reveals central UV emission from the E0 galaxy which could be a typical example of a UV upturn \citep{Kaviraj2007b} or star-forming ellipticals \citep{Elmegreenetal2005a,Dhiwaretal2023,Pandeyetal2024}. Our next example is that of an interacting spiral galaxy at $z \sim 1$ (confirmed by HST grism G141 data). Both HST/F606W and JWST NIRCam/F277W data suggest both spirals could be in the process of merging. The observed N242W band probes rest-frame $\lambda =960 - 1460$\AA~ emission from this merging pair - a powerful way to investigate star-formation at this wavelength. The edge-on galaxy at the bottom of this panel is at $z=0.47$ and the N242W filter probes the stellar emission at rest-frame far-UV. Our third example is that of a barred spiral galaxy at $z\sim 0.21$. The spiral structure appears much clearly in the JWST/NIRCam/F277W observation. The visual comparison suggests that the F154W emission (rest-frame $\lambda = 1074 - 1487$\AA) is present all over the galaxy including the central bar region, indicating the formation of young stars in the bar region - a case similar to what is observed in Malin 1 \citep{Sahaetal2021} in the F154W band. It is worth mentioning here that the size of the galaxy in JWST is large enough that UVIT can resolve it. 

Finally, we handpick a few candidate LyC leakers from our deep observation. Note that the Lyman limit for F154W passband is at $z > 0.975$ while that for N242W band is at $z > 2.35$. Figure~\ref{fig:LyC} shows two candidate leakers based on the F154W data at $z \sim 0.99$ and 1.40. Interestingly, both the galaxies appear extended in HST/F160W band; the morphology of an edge-on galaxy and apparently contamination free. The last panel of Figure~\ref{fig:LyC} shows a candidate LyC leaker at z=2.72 based on our N242W data. At these redshifts, the F154W band will probe a rest-frame wavelength of $600 - 700$~\AA of the ionizing photons \citep[see e.g.,][]{Sahaetal2020, Dhiwaretal2024,Maulicketal2024}, placing UVIT in a unique position to explore the ionizing spectrum of leaking galaxies. All the redshifts presented in this figure are from HST Grism G141 observation \citep{Momchevaetal2016}. Future redshift measurements from JWST/NIRSpec instrument will boost this research activity further. These LyC leakers will be presented in a forthcoming paper.     

\section{Summary and Conclusions}
\label{sec:summary}
We present deep FUV and NUV imaging from UVIT on AstroSat covering the widely studied GOODS South deep field. The presented field, termed AUDF South, is imaged in the F154W and N242W bands with 3$\sigma$ detection limits of 27.2 and 27.7 AB mag respectively, at its deepest. Due to the large PSF, the GALEX GOODS South deep field, although it covers a large sky area, is limited by the confusion limit, which for the FUV and NUV are 25.3 and 24 AB mag, respectively. According to \cite{Xuetal2005}, the fainter boundary of galaxy counts in the GALEX FUV and NUV are 23.8 and 23.6 AB mag respectively which are equivalent to their $3\sigma$ detection limit. Our N242W filter is equivalent to the HST/WFC3/UVIS F225W filter in terms of wavelength coverage. For comparison, the quoted 5$\sigma$ (3$\sigma$) detection limit is 27.8 (28.3) AB mag in the F225W filter in the UVUDF survey \citep{Rafelskietal2015} that covers mostly the HUDF area only. The deep portions of the AUDF South images cover the full HST HUDF \citep{beckwith2006}, XDF \citep{Illingworthetal2013}, UVUDF \citep{Rafelskietal2015}, and a large fraction of the HDUV \citep{Oeschetal2018}, CANDELS \citep{koekemoer2011} image regions. While the shallower portions of the AUDF South images cover the rest of the HDUV and CANDELS image regions and beyond with 3$\sigma$ detection limits of 26.84 mag and 27.34 mag for the F154W and N242W bands respectively.
In addition, the deep UV observation of the AUDF south presented here will greatly benefit from a large number of ongoing/completed spectroscopic surveys e.g., 3D-HST survey \citep{Momchevaetal2016}, MXDF \citep{Baconetal2022}, VLT/VIMOS \citep{LeFevre19} and VANDELS \citep{Pentericcietal2018, McLure18}, JWST/JADES \citep{Eisensteinetal2023} - allowing one to investigate detailed rest-frame UV properties of galaxies with spectroscopy.    
We provide carefully constructed separate catalogs of sources identified in F154W and N242W filters for the AUDF South field. This is complementary to the entire HST-covered GOODS South region, particularly enriching the UV coverage of this deep field at wavelengths bluer (and higher survey depth and spatial resolution) than previously possible with other instrumentation (such as GALEX or HST UV bands). We additionally provide a catalog of CANDELS cross-matched sources separately for the F154W and N242W bands. Some of the key steps for preparing the AUDF South images and subsequent source cataloguing are summarized below:

\begin{enumerate}
    \item Following observations by AstroSat/UVIT, the orbitwise raw data for each image were first reduced and then combined. Two circular fields were observed for each filter with a region of overlap. The overlapping region is termed the deep region and has been observed with exposure times of $\sim$64 and $\sim$62 kilosec in the F154W and N242W filers respectively. 
    \item The combined images were astrometrically aligned using stars observed by GAIA in this field. Careful background measurement was carried out for both images. PSF and completeness estimations were also carried out and presented in detail.
    \item Source detection was carried out on the background-subtracted images in a single mode for the F154W, and using hot and cold modes for the N242W filter. The 3$\sigma$ detection limits are 27.2 and 27.7 AB mag for the deep F154W and N242W images respectively. The presented catalog consists of 13495 and 19374 sources detected above the 3$\sigma$ detection limit.
\end{enumerate}

We expect that the AUDF South source catalog will enable a wide-range of extra-galactic science. A few such science cases that are in preparation by the AUDF team members: identification of a sample of galaxies emitting ionizing photons in the redshift range $z \sim 1 - 3$ and beyond; constraining the UV (rest-frame 1500\AA) luminosity function for $z < 1$; investigating the XUV emission around star-forming galaxies at the intermediate redshift range etc. Covering a unique wavelength range, the AUDF South catalog certainly has paramount legacy value complementing multi-wavelength imaging and spectroscopic data that already exists in the GOODS South field from HST and now JWST \citep{Whittakeretal2019,Eisensteinetal2023b}. AUDF south, being one of the largest far-UV and near-UV survey, will add legacy value to the galaxy science planned using ugrizy photometric observation of the Chandra deep field south by the Vera C. Rubin Observatory Legacy Survey of Space and Time (LSST) under its deep drilling field (DDF) program \citep[][see also the sceince white papers for LSST DDF]{Ivezicetal2019}. More immediately, a large footprint of the AUDF-South will be complemented by the upcoming EUCLID survey \citep{EuclidIetal2022, Euclid_overview2024}, which will cover $\sim 1/3$ of the sky, for photometry where there are no HST or JWST observations at present.

\acknowledgements
This paper uses the data from the AstroSat mission of the Indian Space Research Organisation (ISRO), archived at the Indian Space Science Data Centre (ISSDC). The far- and near-UV observations were carried out by UVIT, which was built in collaboration between IIA, IUCAA, TIFR, ISRO and CSA. KS acknowledges support from the Indian Space Research Organisation (ISRO) funding under project PAO/REF/CP167. This research made use of SExtractor \citep{bertin1996}, Matplotlib \citep{matplotlib2007}, Astropy \citep{astropy2013}, photutils \citep{bradley2020} community-developed core Python packages for Astronomy and SAOImageDS9 \citep{joye2003}.

\software{SExtractor \citep{bertin1996}, SAOImageDS9 \citep{joye2003}, Matplotlib \citep{matplotlib2007}, Astropy \citep{astropy2013}, photutils \citep{bradley2020}}

%

\appendix

\section{Key parameter selection for SExtractor run}
\label{sec:sex_param}

\begin{figure}
\includegraphics[width=0.9\textwidth]{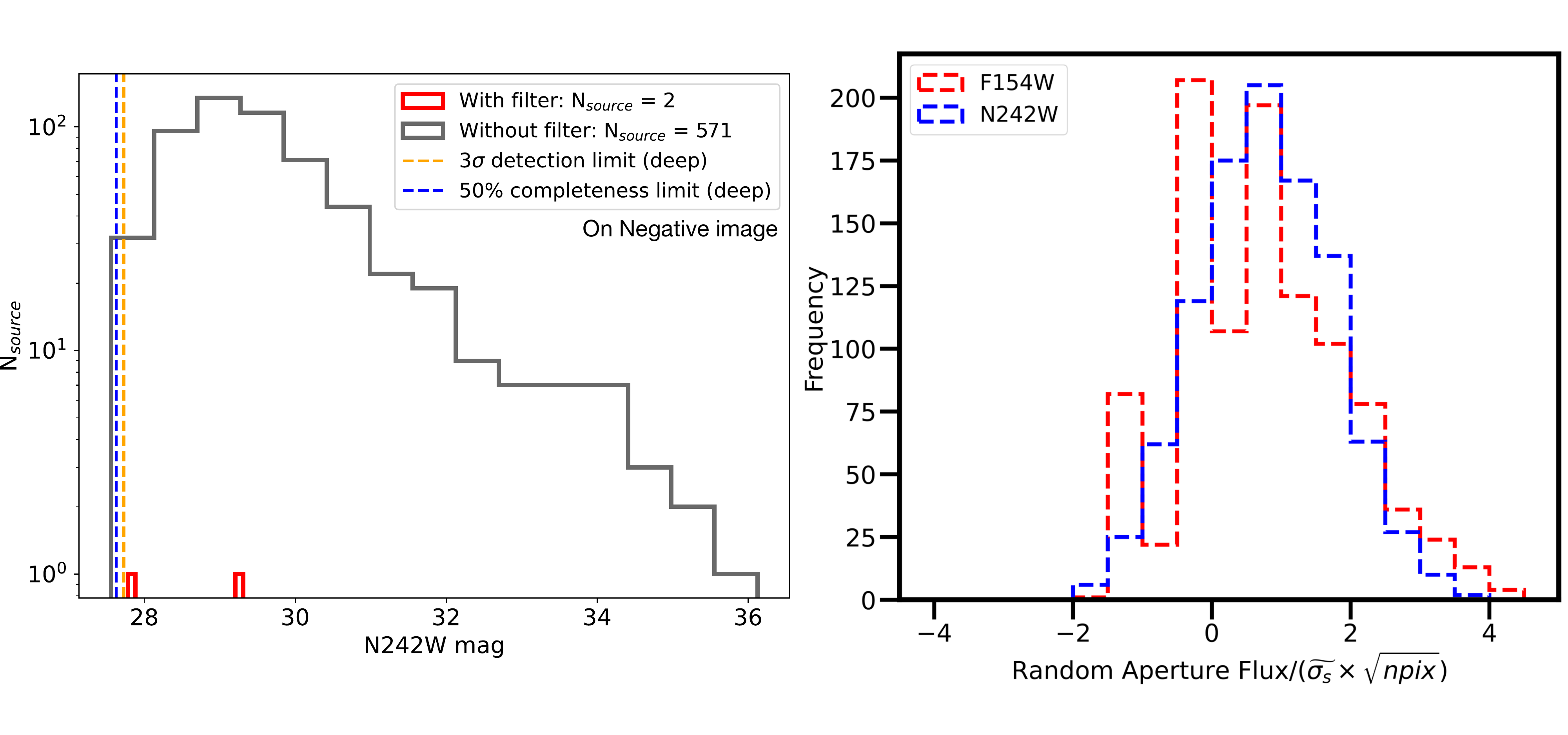}
\caption{\textbf{Left panel:} N242W magnitude of sources detected in the negative image (with and without filter) of the AUDFs field. Two sources detected in our final version are fainter than the $3\sigma$ detection limit. \textbf{Right panel:} Distribution of fluxes within 1000 randomly placed apertures having PSF FWHM diameter, on empty regions, across the F154W (red) and N242W (blue) UVIT images. Note that the flux units are normalized to the sky background rms. $npix$ denotes the number of pixels in the aperture.}
\label{fig:negative_nuv}
\end{figure}

Source detection by the SExtractor algorithm has a strong dependence on the choice of filter kernel, detection MINAREA, detection threshold, and deblending parameters. Choosing an optimized set of these parameters is a daunting task, especially due to issues in simultaneously detecting all the bright and faint sources in a given image. Moreover, the convolution filter (which is a monotonically decreasing function of radius) reduces the contrast between sources, which in turn washes away fainter sources because the automated detection algorithm becomes less effective in separating two nearby sources. Naturally, there is a trade-off between the usage of a filter and no filter. Since our N242W image has a higher source density (compared to GALEX NUV), we decided to first run SExtractor with different Gaussian convolution kernels, including keeping the running of the detection process without a filter as an option. The SExtractor run (with key parameters: MINAREA=5, $Threshold=1.2\sigma$, $DEBLEND\_NTHESH=32$, $DEBLEND\_MINCONT=0.005$) with a Gaussian convolution filter $gauss\_1.5\_3x3.conv$ with an FWHM of 1.5 pixels, on the N242W image resulted in a loss of a large number of sources (more than 50\%) in the neighborhood of the $3\sigma$ detection limit ($\sim 27.7$ AB mag in the deep region of N242W image) compared to the run without a convolution filter. Note that the detection of sources brighter than $27$~AB mag remains unaffected. Several of these detection (i.e., from the run without a filter) fainter than the $3\sigma$ limit might be real sources and would go undetected if we carry out detection using the above-mentioned Gaussian kernel. Increasing the detection threshold and the detection MINAREA might not always help as faint sources would be missed in this case as well. Nevertheless, carrying out detection without a filter might also produce sources (besides, real ones) that are noise peaks. To investigate this further, we run SExtractor with the same parameters (mentioned above) as in the run without a filter but on the negative N242W image. This resulted in about $571$ sources (see left panel of Figure~\ref{fig:negative_nuv}). Although most of these were fainter than the $3\sigma$ detection limit, the detection parameters should be robust against the detection of these undesirable objects. To circumvent this issue, we decided to run SExtractor with a slightly narrower convolution filter $gauss\_1.2\_3x3.conv$ appropriate for the N242W image, where the FWHM of the Gaussian kernel is 1.2 pixels. When SExtractor was run again on the negative N242W image with these revised set of parameters, only two sources fainter than $3\sigma$ detection limit were detected (see Figure~\ref{fig:negative_nuv}). In other words, our parameters were optimized such that it resulted in a null detection of sources brighter than $3\sigma$ detection limit when applied on the negative image.

In the case of the F154W image, we first ran SExtractor (with key parameters: MINAREA=6, $Threshold=1\sigma$, $DEBLEND\_NTHESH=32$, $DEBLEND\_MINCONT=0.0005$) without a filter. This setup led to the detection of a large number of sources that were much fainter than $3 \sigma$ and whose photometric measurements were unreliable. So we decided to use the convolution filter $gauss\_1.5\_3x3.conv$ appropriate for the F154W image. These chosen convolution filters and other optimised parameters (Section~\ref{sec:Sextractor}) resulted in limiting noise peaks to negligible number in our final catalogue, as discussed in the following section.

\section{Random aperture fluxes across the UVIT images}
 In this section, we investigate the distribution of fluxes within apertures placed across `empty' regions in the UVIT images. We define an `empty' region in the context of the segmentation map generated in Section \ref{sec:Sextractor}. Pixels that belong to a source as identified by Source Extractor have positive integer values whereas the rest of the pixels that do not belong to any source are assigned zero value in the segmentation map. Thus apertures placed on empty regions on the UVIT science images have zero flux when placed at the corresponding position in the segmentation images. 
 Note that we use only the hot mode segmentation for the N242W band since in the construction of the combined hot-cold segmentation we get rid of clumpy extensions of objects which would affect our experiment. We randomly place 1000 PSF FWHM diameter-sized apertures on our defined 'empty regions' across the background subtracted F154W and N242W band science images. The distribution of the fluxes is presented in the right panel of Figure \ref{fig:negative_nuv}. We use normalized flux units in terms of the sky background rms $\widetilde{\sigma_\textrm{s}}$. 
The distribution in both the bands is positively skewed with a mean of $\sim 0.8\widetilde{\sigma_\textrm{s}}$. For conservative source detection parameters a positive skew in the mean may be expected, since, while we ensure the apertures do not intersect any source pixels, the experiment still allows for the random apertures to be placed in regions where pixel values lie above the background mean but do not meet the required source detection parameters. Such regions maybe found near extended bright sources. Theoretically, for pure background fluctuations, we should recover a value of 1 from the standard deviation of the distributions in the right panel of Figure \ref{fig:negative_nuv}. However, we report an overestimation in the case of the F154W, with a recovered standard deviation value of 1.2. For the N242W band, the standard deviation of the empty aperture fluxes is 0.97. The number of apertures with fluxes greater than $3\widetilde{\sigma_\textrm{s}}$ is 47 (12), out of 1000, for the F154W (N242W) band. The overestimation of the mean and standard deviations of such a distribution has been reported by \cite{Rafelskietal2015} in the Hubble UVUDF. We plan to investigate this in further detail in the future exploration of the UVIT images.

\begin{figure}
\centering
\includegraphics[width=0.85\textwidth]{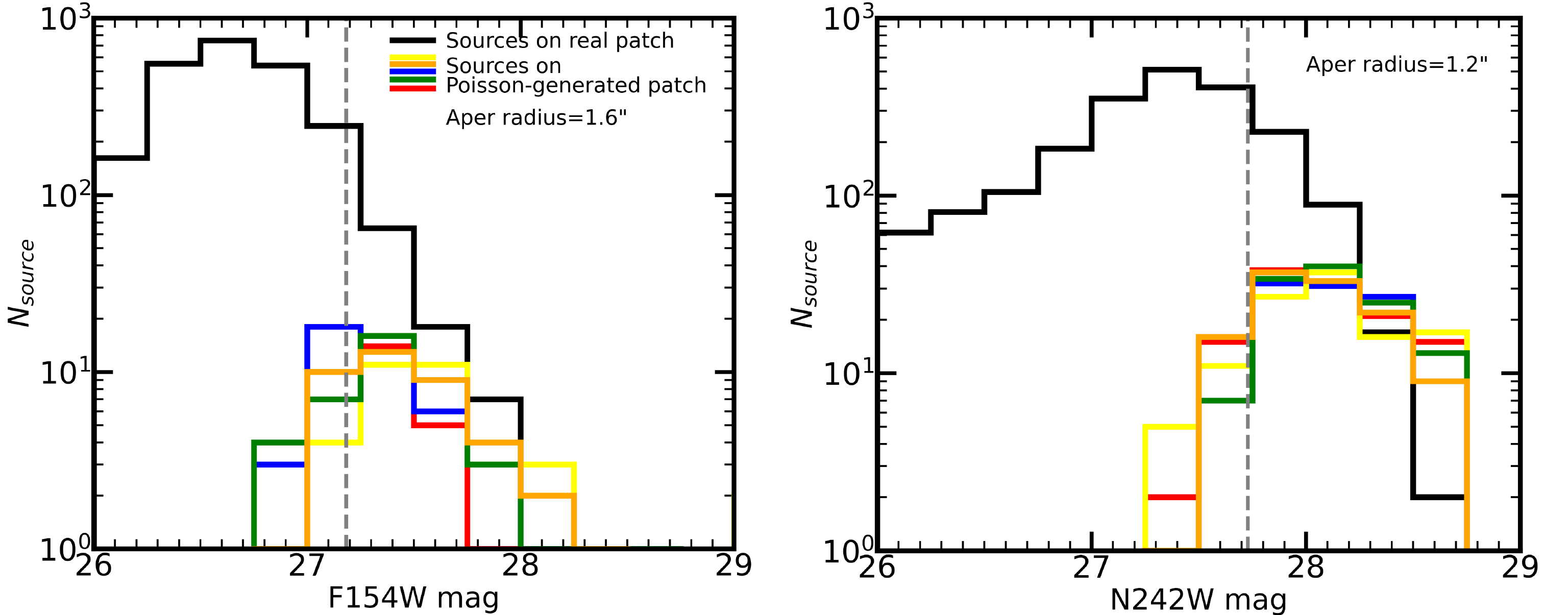}
\caption{Source count distribution (shown are mag$>26$) of detected objects as a function of circular aperture magnitudes in the $4\times4 \: \text{arcmin}^2$ real UVIT (black histogram) field and in the mock Poisson background fields (colored histograms) for F154W (left panel) and N242W (right panel). All noise peaks detected objects resulting from background fluctuations in the Poisson-generated images are fainter than 26 magnitude in both bands. The grey dashed lines indicate the $3\sigma$ detection limits for the deep regions of the N242W and F154W bands. The bin size is 0.25 mag.}
\label{fig:spurious_mag_distbn}
\end{figure}

\begin{table}[]
\caption{The number of detected sources in magnitude bins of width 0.5 from 26.5 to 28.5 AB mag for the experiments carried out in Section \ref{sec:poisson_exp}. The first number in each column indicates the number of sources detected using the detection parameters in Table \ref{tab:sep_table} on the real UVIT image. The number in the parenthesis indicates the average number of sources detected for the particular magnitude bin in the mock Poisson background image.}
\begin{tabular}{|l|l|l|}
\hline
Magnitude bin & F154W    & N242W    \\ \hline
26.5-27       & 1287 (2) & 289 (0)  \\ \hline
27-27.5       & 311 (23) & 866 (2)  \\ \hline
27.5-28       & 25 (11)  & 634 (45) \\ \hline
28-28.5       & 1 (2)    & 106 (57) \\ \hline
\end{tabular}
\label{tab:poisson_exp_table}
\end{table}

\section{Spurious sources, noise peaks detection} 
\label{sec:poisson_exp}

Across any astronomical image there are various artefacts mimicking actual sources. Disentangling these artefacts from real astronomical sources becomes challenging when one is dealing with fainter sources. In some cases, the artefacts could arise due to cosmic ray hit, errors in the data pipeline or imperfection in the instruments and these are called spurious sources. Our pipeline is quite robust to detect hot pixels and flag them off. On the other hand, there may be locations on the image where background noise pixels may cluster together (chance alignment) to possibly mimic a source - called the noise peaks. In other words, noise peaks could arise from pure statistical fluctuations in the background. This becomes particularly important for far-UV and near-UV images where UV emission may arise from a corner of a galaxy and the traditional definition of a real source (where spatial distribution of photons follows a bell-shaped curve, in general) loses its meaning to a large extent. Depending upon the detection parameters used in SExtractor, the number of such noise peak detection arising from these chance alignments may vary. 

In this section, we attempt to identify and quantify spurious detection as well as noise peaks. As mentioned in section~\ref{sec:obs}, during the pipeline processing, all frames hit by cosmic ray showers are identified and removed from the final science-ready images; and thereby reducing the chances of spurious sources occurrence in our final images. Telescope optics have been carefully designed to avoid ghosts and stray light (see the full field UV images in Figure~\ref{fig:cdfsII}). Further, we have visually inspected the whole image to search for any obvious artefacts and found none. In the following, we focus on the estimation of the expected density of noise peak detection in our UVIT images for the adopted Source Extractor parameters (see Section~\ref{sec:catalog}). Before we proceed, we would like to reiterate that through the following experiments, we are considering noise peak detection arising from pure uncorrelated background fluctuations; while in a real image, this marks just one component of the noise and our assumption may not be valid.\\  
For our purpose, we run SExtractor on a randomly generated Poisson background image, which serves as our control image. We take the size of this image to be $576\times576$ pixels corresponding to a $4\times4 \: \text{arcmin}^2$ area. Each pixel in the image is assigned a random value generated using the RANDOM.POISSON function which is a part of the Python module NUMPY. The product of the background and exposure maps (Section \ref{sec:background}) are taken as the input mean for the Poisson distribution function. We also run SExtractor using the same set of detection parameters (see Table \ref{tab:sep_table}) on the real UVIT images having the same size, i.e., a $4\times4 \: \text{arcmin}^2$ field. We note that for detecting sources in the mock N242W Poisson-generated image, we apply only the hot mode detection since bright extended sources are not expected in these images. 

We generate 5 sets of mock images for both the UVIT bands. Since we employ the actual background parameters and exposure maps, our mock background images closely resemble actual background fields. The source counts of detected objects up to 28.5 magnitude are presented in Table \ref{tab:poisson_exp_table}. We do not detect any noise peak brighter than 26.5 magnitude in any of the Poisson-generated background images (for F154W and N242W bands). The first value in the columns indicates the number of sources in the magnitude bin detected in the real patch of the sky while the value within the parentheses are the average counts of detected sources in the same magnitude bin over the 5 mock Poisson-generated background images. The magnitudes are computed within apertures of radius 1.6" and 1.2" for the F154W and N242W bands respectively. We find that the expected number of noise peaks that are formed by chance alignment of background pixels that satisfy our SExtractor detection criteria is significant relative to the number of sources in the representative UVIT real images for magnitudes fainter than the 3$\sigma$ limits of the UVIT images. As mentioned in Section \ref{sec:Sextractor}, our final UVIT catalogs contain only detected objects with SNR greater than three, which more or less coincides with objects that are brighter than the 3$\sigma$ detection limits, where the expected number of such detected noise peaks is negligible (with a space density $\sim 2\times 10^{-4}\: \text{arcsec}^{-2}$ for both bands).


\end{document}